\documentclass[twocolumn]{aastex62}
\newcommand{\Mjup}{\mbox{$\rm M_\mathrm{Jup}$}}
\newcommand{\Msun}{\mbox{$\rm M_{\odot}$}}
\newcommand{\Rsun}{\mbox{$\rm R_{\odot}$}}
\newcommand{\Lsun}{\mbox{$\rm L_{\odot}$}}
\newcommand{\e}[1]{\times 10^{#1}}
\shorttitle{Dynamical Mass of HR 8799}
\shortauthors{Sepulveda \& Bowler}
\begin{document}
\correspondingauthor{Aldo G. Sepulveda}
\email{aldo.sepulveda@hawaii.edu}
\author[0000-0002-8621-2682]{Aldo G. Sepulveda}
\altaffiliation{NSF Graduate Research Fellow}
\affiliation{Institute for Astronomy, University of Hawai'i at M\={a}noa, 2680 Woodlawn Drive, Honolulu, HI 96822, USA}
\affiliation{Department of Physics \& Astronomy, The University of Texas at San Antonio,
1 UTSA Circle
San Antonio, TX, 78249, USA}
\affiliation{Department of Astronomy, The University of Texas at Austin,
2515 Speedway Blvd. Stop C1400 
Austin, TX 78712, USA}
\author[0000-0003-2649-2288]{Brendan P. Bowler}
\affiliation{Department of Astronomy, The University of Texas at Austin,
2515 Speedway Blvd. Stop C1400 
Austin, TX 78712, USA}

\title{Dynamical Mass of the Exoplanet Host Star HR 8799}
\begin{abstract}
HR~8799 is a young A5/F0 star hosting four directly imaged giant planets at wide separations ($\sim$16--78~au) which are undergoing orbital motion and have been continuously monitored with adaptive optics imaging since their discovery over a decade ago. We present a dynamical mass of HR~8799 using 130 epochs of relative astrometry of its planets, which include both published measurements and new medium-band 3.1~$\mu$m observations that we acquired with NIRC2 at Keck Observatory. For the purpose of measuring the host star mass, each orbiting planet is treated as a massless particle and is fit with a Keplerian orbit using Markov chain Monte Carlo. We then use a Bayesian framework to combine each independent total mass measurement into a cumulative dynamical mass using all four planets. The dynamical mass of HR 8799 is 1.47$^{+0.12}_{-0.17}$~\Msun\ assuming a uniform stellar mass prior, or 1.46$^{+0.11}_{-0.15}$~\Msun\ with a weakly informative prior based on spectroscopy. There is a strong covariance between the planets' eccentricities and the total system mass; when the constraint is limited to low eccentricity solutions of $e<0.1$, which is motivated by dynamical stability, our mass measurement improves to 1.43$^{+0.06}_{-0.07}$~\Msun. Our dynamical mass and other fundamental measured parameters of HR~8799 together with MESA Isochrones \& Stellar Tracks grids yields a bulk metallicity most consistent with [Fe/H]$\sim$ --0.25--0.00~dex and an age of 10--23~Myr for the system. This implies hot start masses of 2.7--4.9~\Mjup\ for HR 8799~b and 4.1--7.0~\Mjup\ for HR~8799~c, d, and e, assuming they formed at the same time as the host star.
\end{abstract}

\section{Introduction} \label{sec:intro}
HR 8799 is a A5/F0 V star \citep{grayandkaye1999} that is most well known for hosting four exoplanets that have been directly imaged within a cold extended debris disk and beyond a warmer interior dust belt (\citealt{marios2008,marios2010,Su+2009,boothetal2016,wilner2018,Faramaz+2021}; see also review by \citealt{2018haex}). Among directly imaged systems, HR~8799 stands out as an unusual case of protoplanetary disk evolution due to its multiple planets orbiting at wide separations. In addition to HR~8799, the only other currently known systems with multiple imaged exoplanets are the young stars PDS 70, which hosts two accreting protoplanets within its transition disk \citep{Keppler+2018,Haffert+2019,Mesa+2019}, TYC 8998-760-1, which hosts two recently detected giant planets at wide separations \citep[][]{Bohn+2020a,Bohn+2020b}, and $\beta$ Pictoris, which harbors two giant planets orbiting within 10 au \citep{Lagrange+2009,Lagrange+2010,Lagrange+2019,Nowak+2020}. HR 8799 itself is also unusual with respect to other A stars: it is a $\lambda$ Bo\"{o}tis star with a peculiar surface abundance pattern \citep{grayandkaye1999,Sadakane2006,moya2010b,Murphy+2015} and a $\gamma$  Dor pulsator \citep{Rodriguez&Zerbi1995,zerbi1999,Kaye+1999,moyaetal2010a}. The stellar radius of HR 8799 has been directly measured \citep[][]{bainesetal2012}; however, its mass, age, and bulk composition are either poorly constrained or inferred from models.

Continuous monitoring and ``pre-discovery" observations of the HR 8799 planets have provided a baseline of over two decades of relative astrometry between the companions and host star. Prior studies have constrained the orbits of the companions \citep[e.g.,][]{soummer2011,Currie+2012,esposito2013,gozdziewski2014,pueyo2015,mairetal2015,zurloetal2016,konopacky2016,Wertz+2017,Wang+2018,GRAVITY+2019}, but the host star mass has been assumed to be fixed at $\sim$1.5~\Msun\ in the majority of these analyses based on estimates from surface gravity measurements and high-mass evolutionary models \citep[e.g.,][]{grayandkaye1999,moyaetal2010a,bainesetal2012}.
For example, \citet{mairetal2015}, \citet{zurloetal2016} and \citet{Wertz+2017} adopt a mass of 1.51 \Msun\ based on the results of \citet{bainesetal2012}, who inferred the mass by comparing their stellar radius measurement and effective temperature constraint to high-mass stellar evolution grids. Similarly, \citet{konopacky2016} and \citet{Wang+2018} use a mass prior of 1.52$\pm0.15$ \Msun\ for their orbit fitting analysis, which is also based on the results of \citet{bainesetal2012} coupled with an additional uncertainty to account for model systematics. Dynamical mass measurements, on the other hand, are ``gold standards" because they rely only on Kepler's laws (e.g., \citealt{hillenbrand2004}, see also review by \citealt{Serenelli+2020}). While host star dynamical masses have been measured in other systems with a directly imaged substellar companion \citep[e.g., $\beta$ Pic,][]{Nielsen2014,Wang2016,Dupuy+2019,Gravity+2020,Nielsen+2020,Lagrange+2020,Vandal+2020,Brandt+2021}, it is more often the case that the modest orbital coverage from long-period planets cannot admit a well constrained dynamical mass \citep[e.g.,][]{Bowler+2020}. 

The precise age of HR 8799 remains unsettled but is important because it is necessary to infer mass estimates of the orbiting companions using substellar evolutionary models \citep[e.g.,][]{marleyetal2007,fortneyetal2008,marleau&cumming2014}. If HR 8799 is a member of the Columba Association \citep{Doyon2010,zuckerman2011,maloetal2013} then its age is $42^{+6}_{-4}$ Myr \citep{Bell+2015}. However, \citet{hinzetal2010} calculated the \textit{UVW} space motion of HR 8799 using revised Hipparcos astrometry and called into question whether it is an unambiguous kinematic member of this group. They used their revised space motion together with modeling of the epicyclic orbit of HR 8799 and found that its distance from the center of Columba is implausibly large. \citet{moyaetal2010a} conducted an asteroseismic analysis of HR 8799 and estimated plausible ages as young as 26 Myr and as old as $\sim$1.6 Gyr; however, their interpretation depends on the unknown inclination of the host star\footnote{\citet[][]{Wright+2011} conducted a spectroscopic asteroseismic analysis and found the stellar inclination to be $i_{\star} \gtrsim 40^{\circ}$. However, asteroseismic studies of HR 8799 are generally difficult to carry out due in part to the complex nature of its pulsation frequencies \citep[e.g.,][]{Sodor+2014}.}. \citet{Faramaz+2021} reevaluated the moving group membership status of HR 8799 by calculating the \textit{UVW} space motion with the latest \textit{Gaia} EDR3 astrometry \citep{GaiaEDR3} and using the BANYAN $\Sigma$ tool \citep{BANYAN}. When using the stellar radial velocity constraints derived from \citet{Wang+2018RV} and \citet{Ruffio+2019}, they found that HR 8799 is more likely to be a field star rather than a member of Columba.  \citet{Brandt+2021DMe} recently presented a dynamical mass for the planet HR~8799~e and used it to derive a hot-start cooling age of $42^{+24}_{-16}$ Myr.

In this study, we independently constrain the dynamical mass of HR 8799 by fitting orbits to relative astrometry of each of the four companions utilizing \emph{all} previously published astrometry together with a new epoch we obtained of planets b, c, and d using Keck/NIRC2 in 2015. We then jointly derive the dynamical mass distribution of HR 8799 in a Bayesian framework by assuming the total mass interior to each orbit is simply the stellar mass ($M_{total} \approx M_{\star}$). HR~8799's dynamical mass, measured radius, bolometric luminosity, and effective temperature together with high-mass stellar evolutionary models offer a way to determine the system age independent of kinematic constraints, which in turn can be used to refine the inferred masses of the four planets using substellar evolution models.

In \S \ref{sec:obs} we describe our new observations of the outer three planets with Keck/NIRC2 and summarize the additional astrometry that we use in our analysis. We detail the orbit fitting procedure and our Bayesian framework to jointly constain the host star mass in \S \ref{sec:analysis}. Results are summarized in \S \ref{sec:results} and we explore the implications of our dynamical mass in \S \ref{sec:discussion}.

\section{Observations} \label{sec:obs}
\begin{figure*}[ht!]
  \includegraphics[trim=2.5cm 4.25cm 5cm 5cm, clip,width=7.1in]{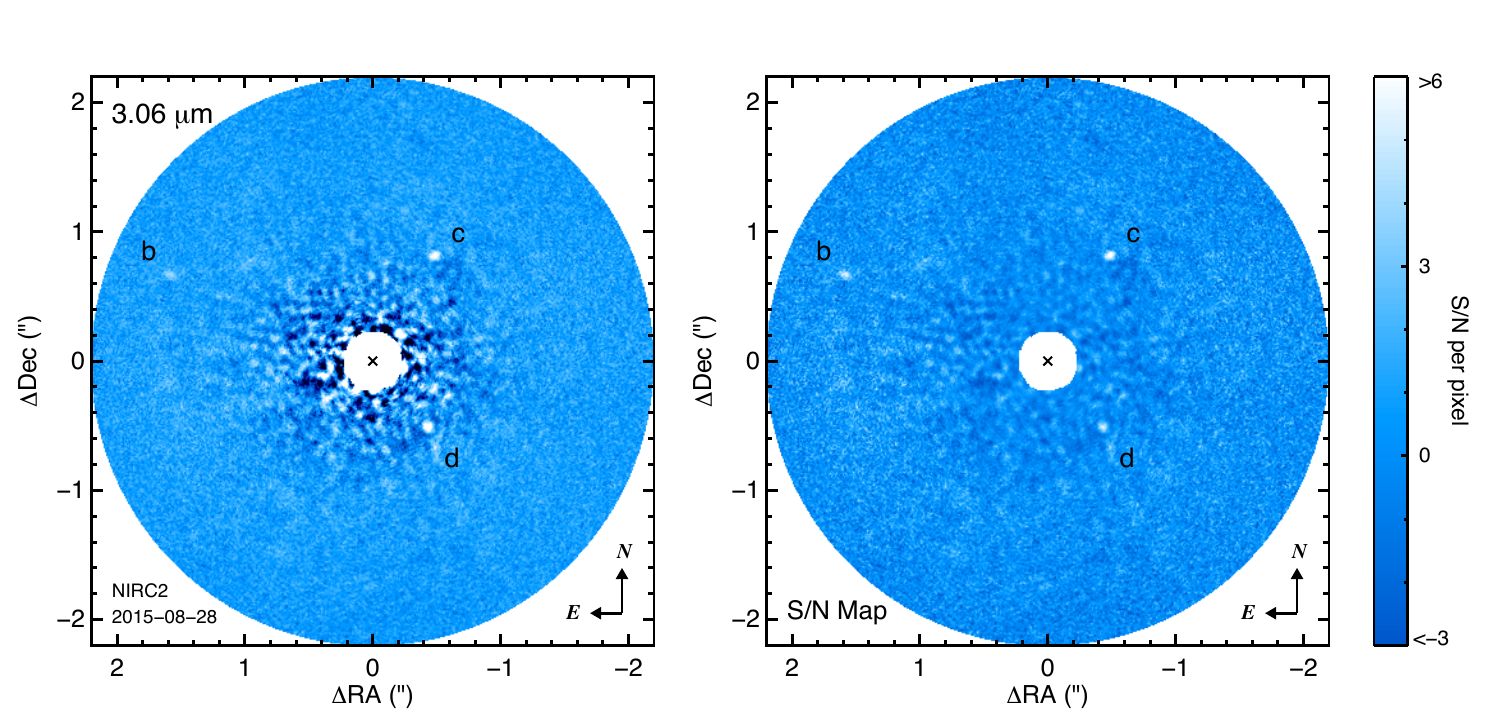}
  \centering
  \caption{NIRC2 3-$\mu$m observation of HR 8799. The left panel shows the processed (PSF-subtracted) image, revealing HR 8799 b, c, d. Planet e is not detected at a significant level in this dataset.  The right panel shows the corresponding S/N map. North is up and east is to the left. \label{fig:keckastrometry}} 
\end{figure*}
\subsection{New Keck/NIRC2 Astrometry}

We observed HR 8799 on UT 2015 August 28 with the NIRC2 near-infrared camera coupled with natural guide star adaptive optics at Keck Observatory (\citealt{Wizinowich:2013dz}). The narrow camera mode was used, providing a plate scale of 9.971$\pm$ 0.004 mas pix$^{-1}$ (\citealt{Service:2016gk}) and a 10$\farcs$2$\times$10$\farcs$2 field of view. To sample a bright region of the planets' emergent spectrum and yield high Strehl ratios, thermal infrared observations were taken with the \emph{H$_2$O ice} filter ($\lambda_\mathrm{central}$ = 3.063~$\mu$m; $\Delta \lambda$ = 0.155 $\mu$m). A total of 119 images were acquired in pupil-tracking (ADI; \citealt{Liu2004,Marois:2006df}) mode over the course of 2.1 hours with HR 8799 centered behind the 400~mas diameter partly-transparent Lyot coronagraph,  amounting to a total field-of-view rotation of 185.5\degr. Each image has an integration time of 10~s and 3 coadds, totaling 3570~s (0.99~hr) of on-source integration time. Throughout transit, HR 8799 passes close to zenith at Mauna Kea so sky rotation is especially fast;  
during this time the sequence was interrupted for about 20 min during which time we obtained nearby 3~$\mu$m sky frames to serve as flats and 15 unsaturated images of the host star (each with an integration time of 0.1~s and 10 coadds) with the mask removed to photometrically calibrate the deeper coronagraphic dataset. Natural seeing in the visible (0.5~$\mu$m) ranged from 0$\farcs$5--0$\farcs$6 as recorded by the CFHT's Differential Image Motion Monitor.

Bad pixels and cosmic rays were corrected using a nearest neighbor averaging algorithm, and each image was flat-fielded using a median-combined and normalized master sky frame. The bright sky background at 3~$\mu$m is removed by subtracting the master sky frame for each image. Images are registered following the description in \citet{Bowler:2015ja}; a 2D elliptical Gaussian is fit to the host star, which is visible behind the coronagraph, and an ADI data cube centered on HR 8799 is assembled. Each image is corrected for optical distortions with the distortion solution from \citet{Service:2016gk}. PSF subtraction is carried out using LOCI (\citealt{Lafreniere:2007bg}) with the following parameter values that control the angular tolerance as well as optimization and subtraction zone sizes and shapes: $W$=7, $N_A$=300, $g$=1.0, $N_{\delta}$=0.5, and $dr$=2.0. Circular masks are placed at the locations of HR 8799 b, c, d, and e to avoid biasing the optimization regions during PSF subtraction.

\begin{deluxetable*}{CCCCCCC}[ht!]
\tablehead{
\colhead{Planet} & \colhead{UT Date} & \colhead{$\rho$} & \colhead{$\theta$} &\colhead{$\Delta$RA} &\colhead{$\Delta$Dec} &\colhead{$\Delta$mag}\\&&(\rm mas)&($^{\circ}$)&($\arcsec$)&($\arcsec$)&
}
\caption{Astrometric and photometric measurements from our NIRC2 observations. \label{tab:astrometry}} 
\startdata
\rm b&2015\ \rm Aug\ 28&1705$\pm$12&67.0$\pm$0.2&1.569$\pm$0.011&0.666$\pm$0.007&12.0$\pm$0.5\\
\rm c&2015\ \rm Aug\ 28&945$\pm$7&329.3$\pm$0.2&-0.482$\pm$0.005&0.813$\pm$0.006&10.9$\pm$0.3\\
\rm d&2015\ \rm Aug\ 28&671$\pm$16&220.5$\pm$0.3&-0.436$\pm$0.011&-0.510$\pm$0.012&10.9$\pm$0.5\\
\enddata
\end{deluxetable*} 

\begin{figure}[t!]
  \includegraphics[trim=0.75cm 2cm 0cm 3cm, clip,width=3.3in]{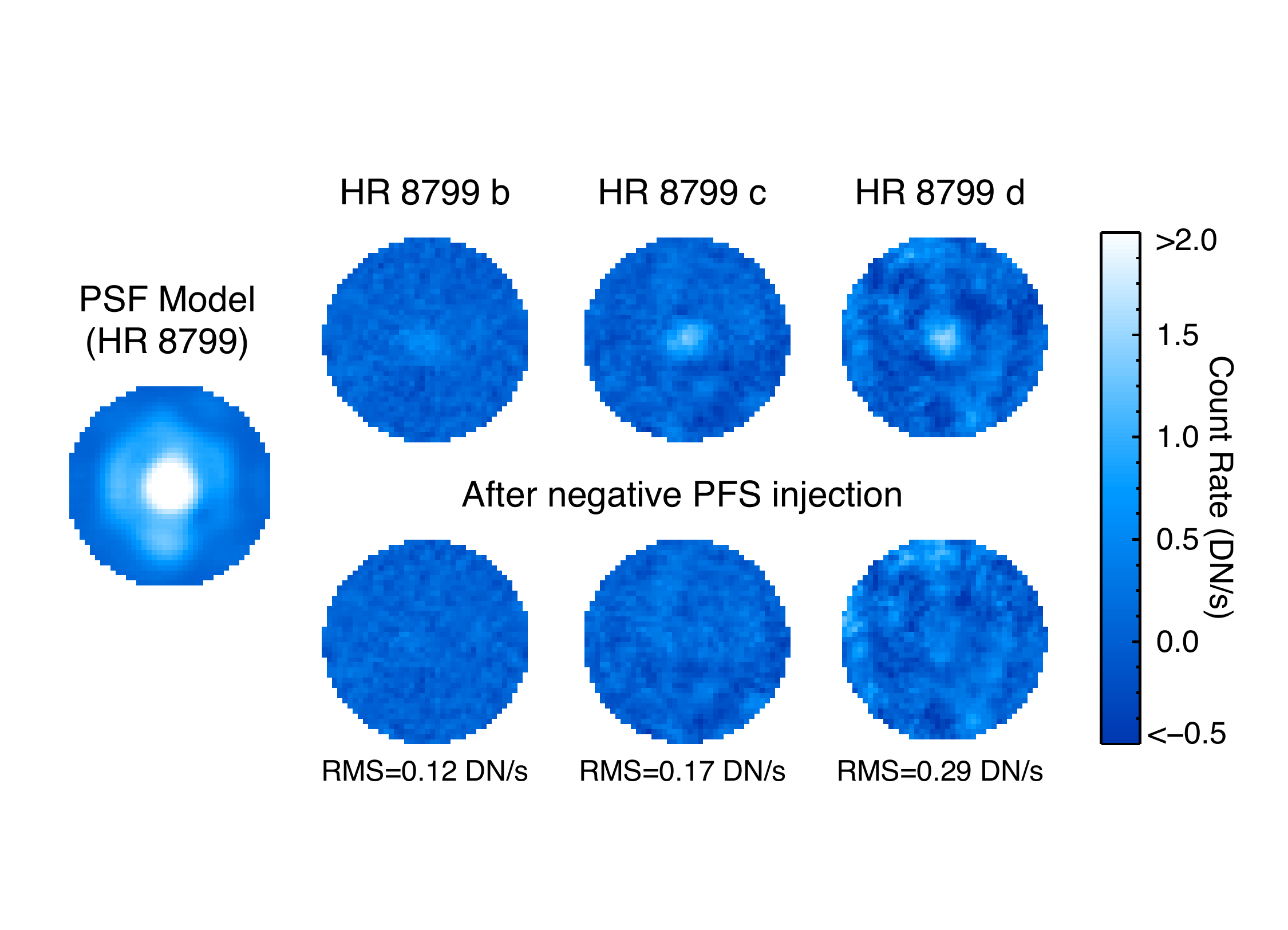}
  \centering
  \caption{Cutouts showing our negative PSF injection strategy for astrometry and relative photometry.  The unsaturated image of HR 8799 (left) is used as a PSF model.  After optimizing the model position and amplitude, planets b, c, and d (top) are removed from the final processed image (bottom) while preserving the underlying noise structure. \label{fig:hr8799_psfinjection}} 
\end{figure}

The final processed image and signal-to-noise ratio (S/N) map are shown in Figure~\ref{fig:keckastrometry}. Planets b, c, and d are clearly visible, but planet e is not evident at a significant level in the processed frame. S/Ns for each planet are computed using non-overlapping circular apertures of 6~pix at 36 azimuthal angles centered on HR 8799; we measure S/Ns of 9.8, 7.3, and 7.2 for planets b, c, and d, respectively.  Astrometry and relative photometry for each source is calculated using the negative PSF injection approach described in \citet{Bowler:2018gy}. The median-combined unsaturated image of HR 8799 is used as the PSF model for each planet. The PSF model is inserted in the raw images with a starting guess position and (negative) amplitude. PSF subtraction is carried out in the same fashion as for our final processed image, and the RMS of the residuals at the location of the planet is used to assess the quality of subtraction. The \texttt{amoeba} downhill simplex algorithm \citep{Nelder:1965tk} is used to find the best-fitting separation, P.A., and flux ratio of each planet (Figure~\ref{fig:hr8799_psfinjection}). The uncertainties in these quantities are computed using the standard deviation of the last ten iterations of the \texttt{amoeba} algorithm as it settles in its final value that minimizes the RMS at a specified tolerance level.  Based on results from \citet{Bowler:2018gy}, this method for estimating errors is comparable to values inferred by repeatedly injecting PSF templates into the raw data at various azimuthal angles and recovering them with the negative PSF injection approach. Results are reported in Table~\ref{tab:astrometry}.

\subsection{Previous Astrometry}\label{subsec:prevAstrometry}
We make use of all astrometry of the HR 8799 planets published prior to December 2020\footnote{During the course of this analysis, \citet{Biller+2021} and \citet{Wahhaj+2021} presented new astrometry from VLT/SPHERE-IRDIS. While these data were not included in this study, continued astrometric monitoring will be essential for future studies of HR~8799.} in our orbit fitting analysis in addition to our new epoch from 2015. Observations of this system have been taken with a wide range of telescopes and instruments, including \textit{HST}, Subaru, Keck, Gemini, MMT, VLT, Palomar, and LBT. Astrometry published prior to 2016 are presented in \citet[][Table 4 in their Appendix]{bowler2016}\footnote{We also include the 2010.55 and 2012.83 Keck/NIRC2 epochs presented in \citet{Currie+2014}, which were inadvertently omitted in the \citet{bowler2016} compilation.}. Additionally, we incorporate published astrometry obtained with Gemini/GPI \citep{Wang+2018} and with VLTI/GRAVITY \citep{GRAVITY+2019}. For astrometry in $\Delta$RA and $\Delta$Dec, we convert to separation and position angle and propagate the uncertainties in a Monte Carlo fashion. For multiple reductions of the same dataset, or where more than one dataset exists for the same epoch, we use the measurements that appear to have the most realistic astrometric errors. Specifically, for epoch 1998.83 we adopt astrometry from \citet{soummer2011} over \citet{Lafreniere+2009}; \citet{Currie+2011} over \citet{hinzetal2010} for epochs 2008.89 and 2009.70; \citet{Currie+2014} over \citet{marios2010} for epoch 2010.55; \citet{konopacky2016} over \citet{marios2008}, \citet{Metchev+2009}, \citet{marios2010}, \citet{Galicher+2011}, and \citet{Currie+2012} for epochs 2007.58, 2007.81, 2008.72, 2009.58, 2009.83, 2010.53, and 2010.83, respectively; \citet{Wertz+2017} over \citet{zurloetal2016} and \citet{Apai+2016} for epoch 2014.93; and \citet{DeRosa+2020} over \citet{Wang+2018} for epochs 2013.88, 2014.70, and 2016.72, respectively. Including our observations from 2015, this results in 35 epochs for planets b (1998.8-2015.7), 36 epochs for planets c (1998.8-2016.7), 35 epochs for planets d (1998.8-2016.7), and 24 epochs for planets e (2009.6-2018.7).

\section{Orbit Fits and Statistical Framework} \label{sec:analysis}
\begin{figure*}[ht!]
  \includegraphics[trim=3.5cm 0cm 3.5cm 0cm, clip, width=7.1in]{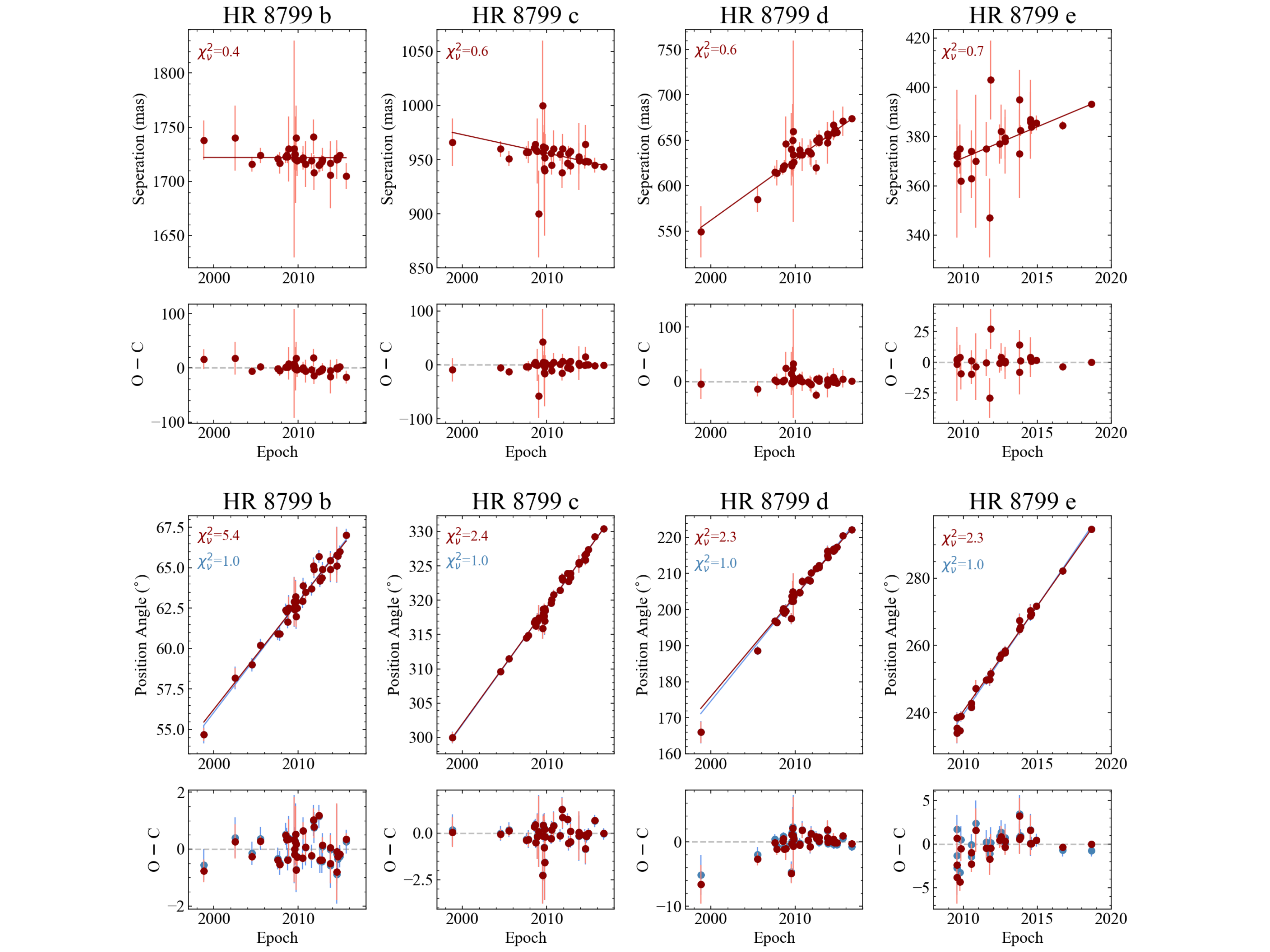}
  \centering
  \caption{Relative astrometry used in this study as a function of time. The best fitting linear models for $\rho (t)$ and $\theta (t)$ are used to estimate the error inflation and calculate the respective $\chi_{\nu} ^2$ values. Raw uncertainties are denoted in red. Position angles have an additional systematic uncertainty term added in quadrature (see \S \ref{subsec:syserr}), denoted in blue along with corresponding revised linear regression. Uncertainties associated with the separation measurements are left unchanged because the reduced $\chi_{\nu} ^2$ value of a linear fit to the raw values is $<1.0$ for all four planets.\label{fig:jitterCombined}} 
\end{figure*}

\begin{figure*}[ht!]
  \includegraphics[trim=0cm 9.5cm 0cm 0cm, clip, width=7.1in]{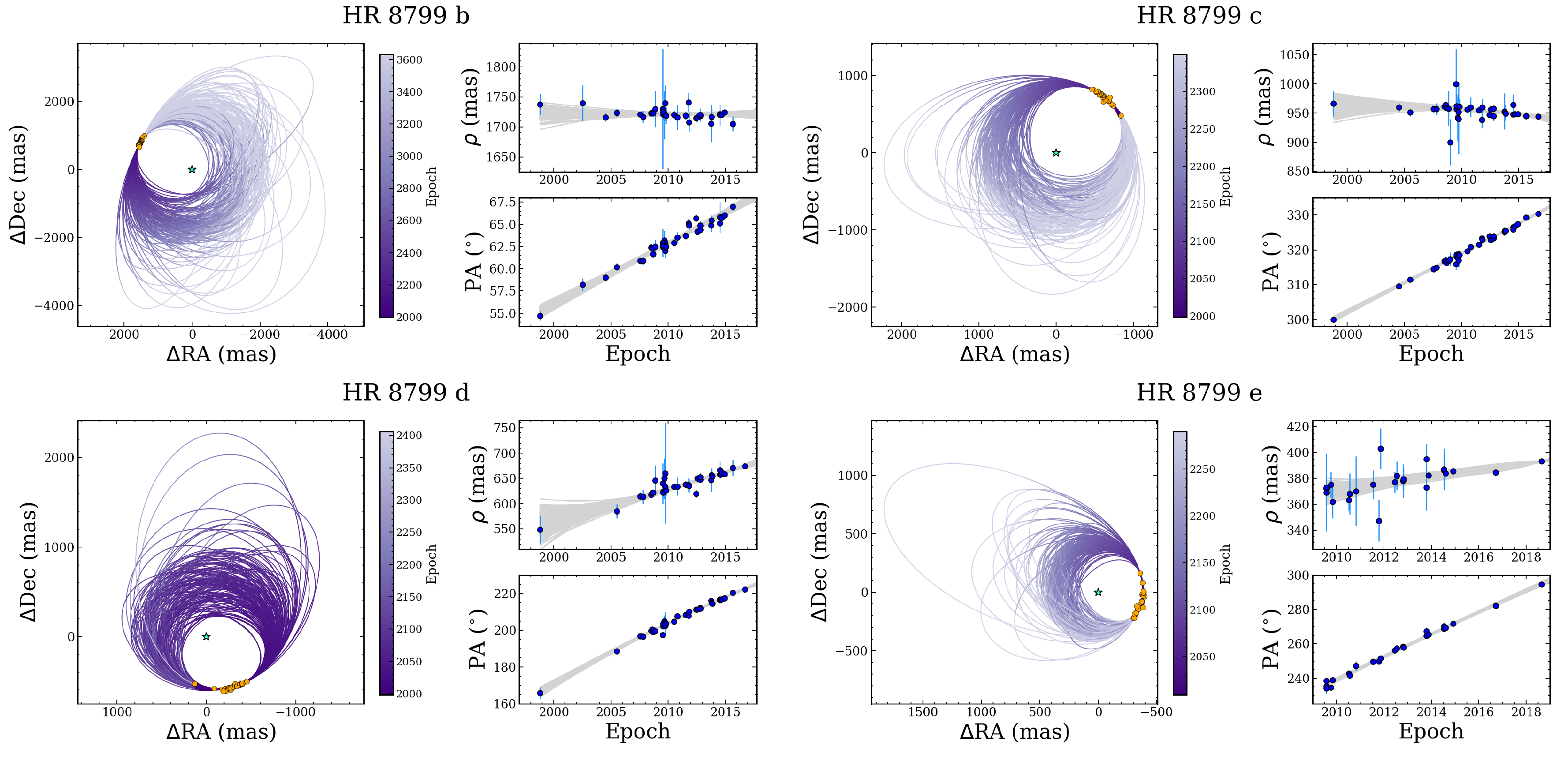}
  \centering
  \caption{Orbit fits for HR~8799 b, c, d, and e using \texttt{orbitize!}. Each panel shows 150 orbits drawn from the converged MCMC chains as described in \S\ref{sec:orbitfitting} together with the measured astrometry. For each planet, the left panel displays the orbits as they appear on-sky and the right panels compare the same orbits to the astrometry. \label{fig:quadorbitsamples}} 
\end{figure*}
\subsection{Accounting for Potential Systematic Errors}\label{subsec:syserr}
Using all available astrometry in the orbit fitting could introduce systematic errors from a variety of sources, including the use of differently calibrated instruments, PSF-subtraction algorithms, and strategies to measure relative astrometry. One approach to avoid this problem is to adopt a uniform dataset from a small number of well-calibrated instruments. For example, \citet{konopacky2016} used data obtained solely from the NIRC2 camera at Keck Observatory to create self-consistent measurements for their orbit fitting. Several other recent studies adopted similar strategies \citep[e.g.,][]{Wang+2018,GRAVITY+2019}. However, while this approach may mitigate systematic errors, it also limits the total time baseline and number of astrometric epochs for the orbit fits. 

Here we use all available astrometry of the HR 8799 planets as of December 2020 with the goal of maximizing the information content and time baseline of the data. Given that potential systematic uncertainties may affect our dynamical mass measurement, we consider an additional uncertainty term that we refer to as astrometric ``jitter" ($\sigma_{\rm jit}$) to be added in quadrature to each set of raw astrometric errors following the approach in \citet{Bowler+2020}. For each planet, we assume a linear model for both separation ($\rho$) and position angle ($\theta$) as a function of time because the astrometric time baseline relative to the orbital period is small ($\lesssim$ 0.16 for all planets). We adopt a reduced $\chi ^2$ value ($\chi_{\nu} ^2 \equiv \chi ^2 /\nu$, where $\nu$ is the number of degrees of freedom) as a metric to assess reliability of the quoted uncertainties. The $\chi_{\nu} ^2$ values for all four planets' separations as a function of time are $<1.0$ (0.37, 0.59, 0.61, and 0.69 for HR 8799 b, c, d, and e, respectively), indicating that the raw uncertainties appear to be reasonable. These are left unchanged in this study. However, all four planets' position angles as a function of time have $\chi_{\nu} ^2 > 1.0$ (5.4, 2.4, 2.3, and 2.3 for HR 8799 b, c, d, and e, respectively), suggesting that the quoted uncertainties are underestimated or that systematic errors are present. The PA jitter term ($\sigma_{\rm jit,PA}$) for each planet is found by iteratively increasing a starting value of 0 in steps of $10^{-5 \circ}$ until $\chi_{\nu} ^2$ reaches unity. This results in $\sigma_{\rm jit,PA}$ values of 0.37$^{\circ}$, 0.26$^{\circ}$, 0.62$^{\circ}$, and 0.69$^{\circ}$ for HR 8799 b, c, d and e, respectively. These are added in quadrature to the original uncertainties: $\sigma_{\rm total,PA}^{2}=\sigma_{\rm jit,PA}^{2}+\sigma_{\rm PA}^{2}$. The original and adjusted astrometry are displayed in Figure \ref{fig:jitterCombined}.
\begin{figure*}[t!]
  \includegraphics[trim=1cm 0.5cm 1cm 1cm, clip,width=7.1in]{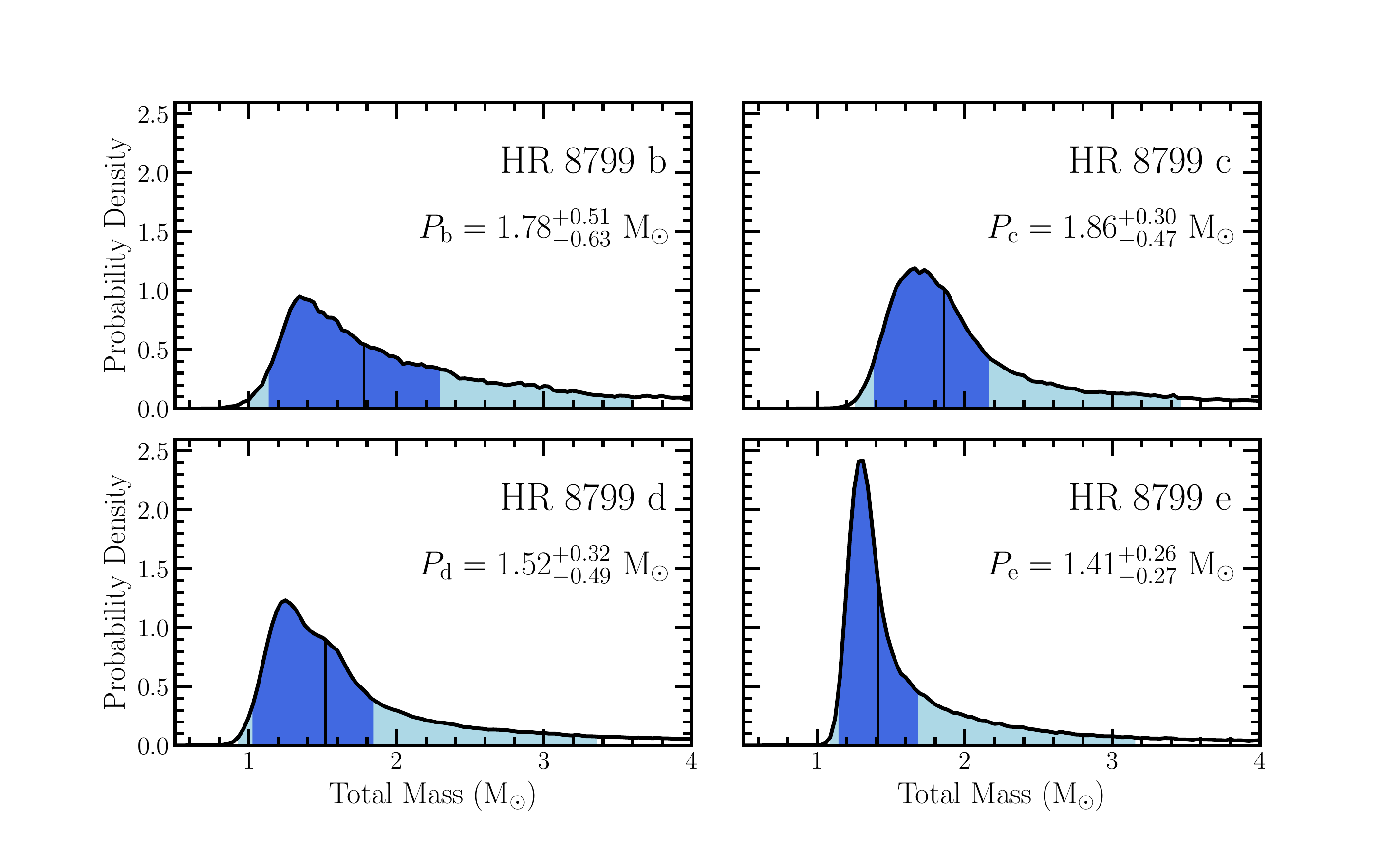}
  \centering
  \caption{Posterior probability distributions for the total mass of HR 8799 based on orbit fitting for each individual planet as described in \S \ref{sec:orbitfitting}. Dark blue regions represent the 68\% HCI and light blue regions represent the 95\% HCI. Median values are denoted with a black bar. \label{fig:interp}} 
\end{figure*}
\subsection{Orbit Fitting}\label{sec:orbitfitting}
We fit Keplerian orbits to the astrometric data of each planet using the \texttt{orbitize!} \citep{orbitize} Python package. Here the \texttt{ptemcee} \citep{foremanmackey13,ptemcee} branch is utilized, which samples the posterior distributions of orbital elements with the parallel-tempered affine-invariant ensemble sampler implementation of Markov chain Monte Carlo \citep[MCMC;][]{goodmanWeare2010}. In the Keplerian model used in \texttt{orbitize!}\footnote{For a more complete description, see \citet{orbitize} and  \url{https://orbitize.readthedocs.io}}, the following parameters are varied directly when fitting with relative astrometry: $a$ (semi-major axis), $e$ (eccentricity), $i$ (inclination), $\omega$ (argument of periapsis), $\Omega$ (position angle of the ascending node), $\tau$ (time of last periapsis), $\pi$ (parallax), and $M_{tot}$ (total mass of the system). We treat each planet independently, corresponding to four separate orbital constraints.

A Gaussian prior of 24.46 mas with a standard deviation of 0.05 mas from \textit{Gaia} EDR3 \citep{gaiamission,GaiaEDR3} is used for $\pi$. A log-uniform prior distribution is adopted for $a$ (i.e., $P(a)\propto a^{-1}$) with a range of 1 to 150 au. An isotropic inclination distribution is adopted for $i$ such that $P(i)\propto \sin i$ (which spans 0 to $\pi$ rad). Linearly uniform priors are chosen for the remaining free parameters. Their ranges are: 0 to 1 for $e$\footnote{\citet{Bowler+2020} found that the underlying eccentricity distribution for a sample of 27 imaged substellar companions is approximately uniform, which motivates our use of a linearly uniform eccentricity prior.}; 0 to 2$\pi$ rad for both $\omega$ and $\Omega$; 0 to 1 for $\tau$; and 0.5 to 4.0 \Msun\ for $M_{tot}$. To avoid artificially reducing the precision of the final dynamical mass measurement, we make no further assumptions about the nature of the planets' orbits such as coplanarity, resonances, stability, or orbit crossing events. 

Each \texttt{ptemcee} ensemble is initialized with 16 temperatures and 1500 walkers per temperature. Each walker samples from the parameter space 1.2$\e{5}$ times and a burn-in length of the first 6.0$\e{4}$ steps is discarded. To assist their convergence, the walkers are initialized at positions near the most probable values found from conducting preliminary orbit fits. Saving only the samples from the lowest-temperature set of chains and applying a thinning factor of 20 to curtail the effect of correlation results in 4.5$\e{6}$ posterior samples for each orbit fit.

\begin{figure*}[th!]
  \includegraphics[trim=1.25cm 0cm 1.25cm 1.55cm, clip,width=7.1in]{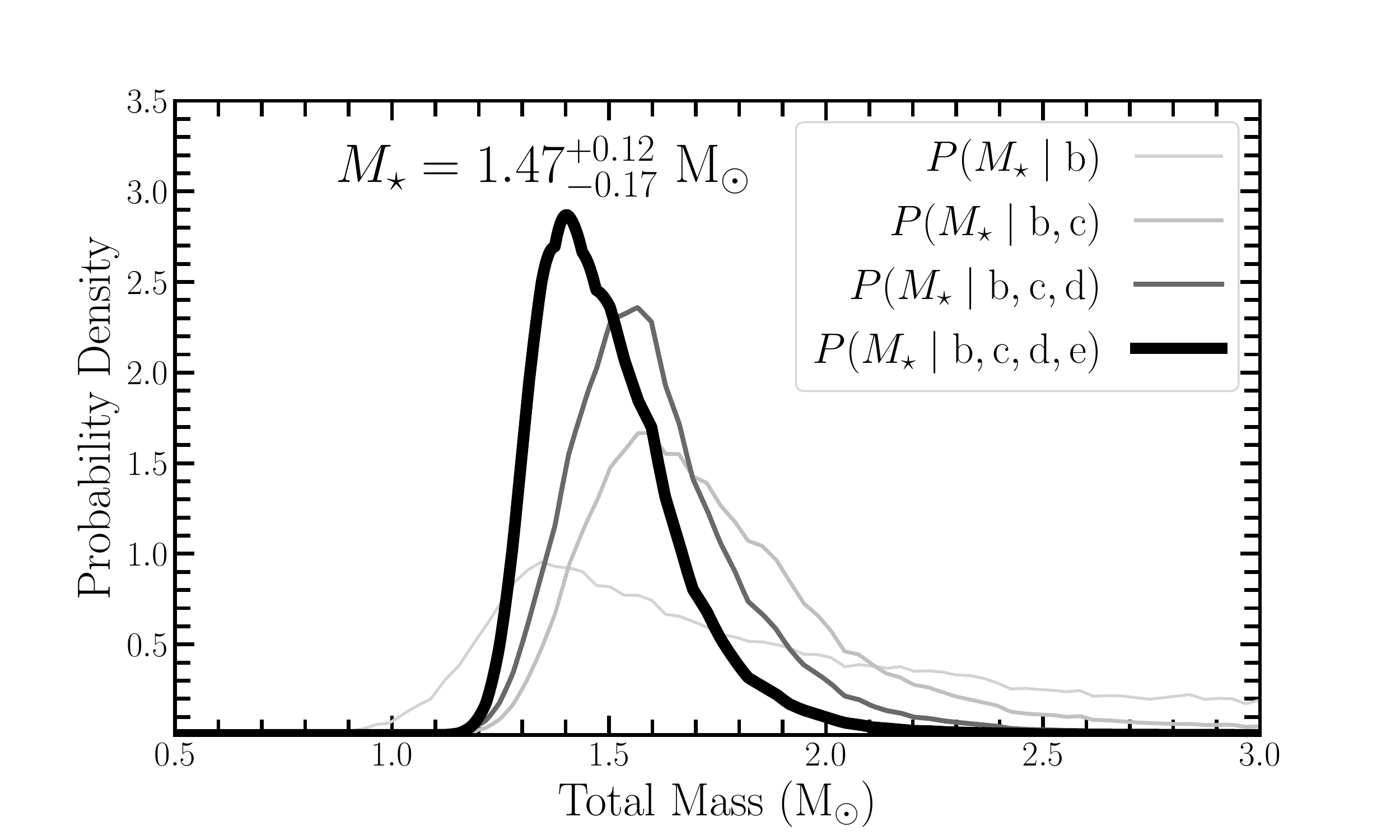}
  \centering
  \caption{Graphical representation of our Bayesian framework to derive the joint dynamical mass posterior distribution for HR 8799. $P(M_{\star} \mid \mathrm{b, c, d, e})$ represents the final joint constraint on the dynamical mass of HR 8799 given the astrometry of all four planets and without any assumptions about coplanarity, stability or orbital resonances. This example uses a uniform prior for $P(M_{\star})$.
  \label{fig:dynamicalmass}} 
\end{figure*}

\subsection{Bayesian Framework to Constrain the Mass of HR 8799}\label{subsec:bayes}
We employ a Bayesian inference framework to derive the joint probability distribution for the host star mass based on our individual orbit fits for each of the four planets. Bayes' theorem naturally updates prior information with new data to improve posterior probabilities:

\begin{equation}\label{bayes}
P (\theta \mid D) \propto P (\theta) \times P (D \mid \theta). 
\end{equation}
Here $P (\theta)$ represents prior knowledge of the parameters $\theta$, $P (D\mid\theta)$ represents the likelihood function of the data $D$ given the model, and $P (\theta\mid D)$ represents the posterior probability of the model parameters given the data. For this study of HR 8799, our orbit fits are carried out for a uniform prior distribution in total (stellar) mass, so each of the four total mass posterior distributions from the independent MCMC fits also represent the likelihood function for the dynamical mass of HR 8799 (to within a constant of proportionality) given the set of astrometry for that planet. This arises by treating the planets as massless ``test" particles such that the total mass of the system is assumed to be confined to the host star ($M_p \ll M_{\star}$, or $M_{\star} \approx M_{tot} $). If the HR 8799 planets have masses $\lesssim 10$ \Mjup, as expected from evolutionary models (e.g., \S\ref{sec:hotstartplanetmasses}), then this approximation is accurate at the $\lesssim 2 \%$ level.

We use the following notation to apply Bayes' Theorem to each of the four independent total mass distributions for planets b, c, d, and e that resulted from orbit fitting:
\begin{eqnarray}
\begin{array}{cc}
P_{\rm b} \equiv  P (\mathrm{b} \mid M_{\star}),\ P_{\rm c} \equiv P (\mathrm{c} \mid M_{\star}),\\
P_{\rm d} \equiv P (\mathrm{d} \mid M_{\star}),\  P_{\rm e} \equiv P (\mathrm{e} \mid M_{\star}). 
\end{array}
\end{eqnarray}
$P_{\rm i}$ is therefore the likelihood function of the astrometry for planet $i$ conditioned on the stellar (total) mass $M_{\star}$. The prior probability for $M_{\star}$, $P(M_{\star})$, is informed by prior knowledge of the stellar mass (e.g., based on spectral type or evolutionary models). For each iteration of planet b, c, d, and e, the host star mass posterior distribution sequentially becomes the new prior at each stage. For planet b, this is $P(M_{\star} \mid \mathrm{b}) \propto P(M_{\star}) \times P_{\rm b}$, for planets b and c this is $P(M_{\star} \mid \mathrm{b, c}) \propto P(M_{\star}) \times P_{\rm b} \times P_{\rm c}$, etc. The jointly constrained mass from all four planets is then:

\begin{equation} \label{eqn:joint}
P(M_{\star} \mid \mathrm{b, c, d, e}) \propto P(M_{\star}) \times P_{\rm b} \times P_{\rm c} \times P_{\rm d} \times P_{\rm e}.
\end{equation}
This setup allows us to only run the orbit fits once for each planet with a uniform mass prior, but then easily adjust the prior $P(M_{\star})$ as desired for the joint mass constraint. 

\section{Results}\label{sec:results}
For each planet, the resulting $M_{tot}$ posterior distributions are linearly interpolated onto a fine grid spanning 0.5--4.0 \Msun\ (Figure \ref{fig:interp}). These are then multiplied together following Equation \ref{eqn:joint} to arrive at the jointly constrained dynamical mass distribution of HR 8799. Given that the individual total mass constraints are non-Gaussian and right skewed, we summarize the results using the posterior distribution median and highest-density credible interval (HCI) metrics.

For the initializing stellar mass prior $P(M_{\star})$, we adopt a linearly uniform distribution to let the astrometry drive the dynamical mass. The final result for this baseline case is shown as Figure \ref{fig:dynamicalmass}. The median and 68\% HCI of the final joint mass posterior is 1.47$^{+0.12}_{-0.17}$ \Msun\footnote{A log-uniform mass prior proportional to $M^{-1}$ instead of a linearly uniform prior for $P(M_{\star})$ results in 1.46$^{+0.12}_{-0.16}$ \Msun. Both constraints are nearly identical.}. The 95\% and 99.7\% HCI span 1.23--1.86 \Msun\ and 1.16--2.32 \Msun, respectively. 

Our choice of a linearly uniform prior may not be especially realistic given that we know HR 8799 is a late-A/early-F star. Our joint stellar mass measurement may be further constrained by using a more informative prior that better captures knowledge of the host star before our experiment takes place while at the same time not being overly restrictive to ensure the astrometry still dominate the posterior. \citet{grayandkaye1999} conducted a spectroscopic analysis of HR~8799 and estimated its mass to be 1.47 $\pm$ 0.30 \Msun\ based on their surface gravity measurement combined with an estimate of the radius. Following the same analysis detailed above but now using this Gaussian prior results in a dynamical mass of 1.46$^{+0.11}_{-0.15}$ \Msun\ for HR~8799. Here the 95\% and 99.7\% HCI span 1.25--1.75 \Msun, and 1.18--1.97 \Msun, respectively. The effect of the informed prior is to further reduce the uncertainties on the final joint stellar mass by $\approx$10\%, which is expected given that this prior is in good agreement with the original dynamical mass (which used a uniform prior). 

While not the focus of this study, we summarize results for the orbital parameters from our fits with cornerplots and a table of best-fit values in Appendix \ref{appendix}. At the 95\% HCI levels, the eccentricities of the planets are constrained to $e\lesssim 0.5$, with the exception of HR~8799~d at $e\lesssim 0.6$. Similarly, at the 68\% HCI levels, the eccentricities of planets b, c, and e are constrained to $e\lesssim 0.3$ while HR~8799~d has $e\lesssim 0.4$. This is consistent with previous studies that also found the allowed eccentricities of HR 8799 d to be marginally greater than those of the other three planets \citep[e.g.,][]{pueyo2015,Wertz+2017}. The most probable inclinations for the four planets span \textit{i}$\approx$25--35$^{\circ}$, which are similar to each other as well as with that of the extended debris disk (\textit{i}$\approx$ 31--33$^{\circ}$, \citealt{wilner2018,Faramaz+2021}). However, the longitude of the ascending node ($\Omega$)\footnote{Orbit fitting with only relative astrometry suffers from a 180$^{\circ}$ degeneracy for both $\Omega$ and $\omega$. In this work, we wrap all posteriors of $\Omega$ and $\omega$ from [0,360]$^{\circ}$ to [0,180]$^{\circ}$. While beyond the scope of this study, radial velocities of the planets can be used to break the $\omega-\Omega$ degeneracy. See, e.g., \citet{Wang+2018RV,Ruffio+2019,Ruffio+2021} and \citet{Wang+2021}.} must be accounted for in addition to $i$ to assess coplanarity. For any two orbital planes, these parameters are related to the true mutual inclination $\psi$ by the geometric relation $\psi=\arccos(\cos i_1 \cos i_2 + \sin i_1 \sin i_2 \cos (\Omega_1 - \Omega_2))$. We place a 95\% confidence upper limit of $\psi< 55-75^{\circ}$ for all permutations of the HR 8799 planet orbital planes. Orbital parameters in this study are generally consistent with results from prior published orbit fits, although our posteriors tend to be larger. This is an expected consequence of our methodology: we solved for the total system mass using agnostic priors, made no constraining assumptions on allowed orbits based on stability, and inflated the astrometric uncertainties to account for potential systematic errors.

\subsection{Low-Eccentricity Scenario}\label{sec:lowe}
The posteriors in Figure \ref{fig:quadcorners} show that there is strong covariance between planet eccentricities and total system mass; higher planet eccentricities imply a substantially larger total mass. To assess how the cumulative dynamical mass might change if near-circular orbits are considered, we reconduct the analysis in an identical manner but we now limit the possible orbit solutions for each planet to $e\leq 0.1$. Although low eccentricities on their own are not necessarily indicative of orbital stability \citep[e.g.,][]{Fabrycky+Murray-Clay2010}, this exercise gives some insight into how the stellar mass constraint depends on eccentricity and what we might expect if long-term orbital stability is taken into account\footnote{For robust explorations and discussion of the dynamical stability of the HR~8799 planets, we defer to recent studies by, e.g., \citet{Wang+2018}, \citet{Gozdziewski+Migaszewski2018,Gozdziewski+Migaszewski2020}, and \citet{Veras+Hinkley2021}.}. The joint dynamical mass in this scenario of near-circular orbits is 1.43$^{+0.06}_{-0.07}$ \Msun, with 95\% and 99.7\% HCI intervals spanning 1.31--1.57 \Msun\ and 1.28--1.65 \Msun, respectively. The smaller uncertainties are expected given the strong correlation between eccentricity and total mass in our baseline orbit fit. Using the informed Gaussian mass prior of 1.47 $\pm$ 0.30 \Msun\ does not significantly influence the joint stellar mass constraint ($M_{\star} = 1.43^{+0.06}_{-0.07}$ \Msun\ in this scenario). The orbit fitting results for this low eccentricity scenario are presented in Appendix \ref{appendix:orbitFitLE}. We conclude that a restricted eccentricity significantly impacts the dynamical mass by reducing the overall uncertainty by a factor of $\approx$2. For this study, however, we adopt the unrestricted solution in which eccentricity is allowed to vary from 0 to 1 to avoid assumptions about mutual planet interactions, which are not yet robustly validated by the astrometry.     

\section{Discussion} \label{sec:discussion}
We fit orbits to all the available astrometry of the HR~8799 planets with minimal dynamical assumptions, then combined the independent total mass constraints into a joint dynamical mass measurement. Table 1 from \citet{bainesetal2012} compiles many of the existing mass estimates of HR 8799, which have been determined with a variety of methods and generally span the range of $\sim$1.2--1.8 \Msun. Our dynamical mass measurement is consistent with all these prior estimates at its 95\% HCI level, including the estimates of 1.513$^{+0.023}_{-0.024}$ \Msun\ and 1.516$^{+0.038}_{-0.024}$ \Msun\ derived by \citet{bainesetal2012} from their measured stellar parameters together with Yonsei-Yale \citep{Yi+2001} evolutionary models. Our dynamical mass measurement thus serves as a reassuring validation of the $\approx$1.5 \Msun\ stellar mass commonly adopted in previous studies of the HR~8799 system.

\citet{Wang+2018} used a self-consistent astrometric dataset from NIRC2 and GPI (55 epochs in total) to simultaneously measure the companion orbital parameters and total mass of HR~8799 in an MCMC framework. In their ``unconstrained" fits, they found a total mass of 1.48$^{+0.05}_{-0.04}$ \Msun\ which, under the approximation $M_{\star} \approx M_{tot}$, can be considered another dynamical mass measurement of HR~8799 to compare with ours. The larger uncertainties in this study are likely influenced by our adoption of a linearly uniform total mass prior compared to their adoption of a well-informed Gaussian prior of 1.52$\pm0.15$ \Msun. Furthermore, by using a dataset only from NIRC2 and GPI without the inflated astrometric uncertainty that we accounted for, their astrometric measurement uncertainties were overall smaller compared to the astrometry used in this study. However, despite some differences in procedure and data used, both results agree well. The dynamical mass presented in this work should be considered a conservative, model-independent measurement that prioritized astrometric baseline over per-epoch precision and instrument uniformity.

\subsection{Age of HR 8799}\label{subsec:stellar}
Several stellar parameters of HR 8799 (radius, luminosity, and effective temperature) were directly constrained by \citet{bainesetal2012} using a Center for High Angular Resolution Astronomy (CHARA) Array stellar angular diameter measurement together with photometry and the parallax from Hipparcos. They found a radius of $ R_{\star}=1.44\pm$0.06 $\Rsun$, a bolometric luminosity of $ L_{\star}=5.05\pm$0.29 $\Lsun$, and an effective temperature of $T_{\rm eff}= 7193 \pm$87 K; these are based on a bolometric flux of $F_{\rm bol}=$ (1.043$\pm$0.012)$\e{-7}$ erg s$^{-1}$ cm$^{-2}$, an angular diameter (accounting for limb darkening) of $\theta_{\rm LD}$ = 0.342$\pm$0.008 mas, and a distance of $D_{\star}=39.4\pm$1.0 pc \citep{vanLeeuwen2007}. \citet{Faramaz+2021} updated and improved the statistical uncertainties for several of these values. They found $L_{\star}=5.441 \pm$ 0.066 $\Lsun$ using the bolometric flux from \citet{bainesetal2012} together with a distance of 40.851$\pm$0.076 pc from \textit{Gaia} EDR3 \citep{GaiaEDR3,Lindegren+2021}. They also updated the radius to $R_{\star}=1.502\pm$0.035 $\Rsun$ when using their \textit{Gaia} EDR3 distance together with the stellar angular diameter from \citet{bainesetal2012}.

The measured mass, luminosity, radius, and effective temperature of HR 8799 (Table \ref{tab:stellarprop}) can be used to constrain its bulk metallicity and age using stellar evolutionary models. We downloaded a custom grid of MESA Isochrones \& Stellar Tracks (MIST) evolutionary models \citep{mesa1,mesa2,mesa3,mesa3.5,mesa4,mesa5}  spanning $M_{\star}$ = 0.5--4.0~\Msun\ ($\Delta M_{\star}$ = 0.01~\Msun) for four metallicities\footnote{Here [Z/H] = [Fe/H], the solar-scaled bulk metallicity.}: [Fe/H] = --0.5, --0.25, 0.0, and +0.25 dex.  A surface angular velocity at 40\% of the critical value is adopted following \citet{mesa5}; however, the effect of rotation and surface mass loss is minimal in this mass regime.
Intermediate-mass stars like HR 8799 follow a meandering trajectory through the HR diagram as they quickly reach the main sequence and then evolve off it.  In doing so their
luminosity and effective temperature will repeatedly rise and fall.  As a result, a given luminosity and temperature can correspond to two or more degenerate ages during the pre- and post-main sequence evolutionary phases.  The additional mass and radius measurements help by limiting the ages consistent with all of these constrained parameters.

\begin{deluxetable}{lcc}
\tablehead{
\colhead{Parameter} & \colhead{Value} & \colhead{References}
}
\caption{Measured Fundamental Stellar Parameters of HR~8799 \label{tab:stellarprop}} 
\startdata
$M_{\star}$ ($\Msun$)&1.47$^{+0.12}_{-0.17}$\tablenotemark{a}&1\\
$R_{\star}$ ($\Rsun$)&1.502$\pm$0.035&2,3\\
$L_{\star}$ ($\Lsun$)&5.441$\pm$0.066&2,3\\
$T_\mathrm{eff}$ (K)&7193$\pm$87\tablenotemark{b}&3\\
\enddata
\tablereferences{(1) This work (assumes a uniform stellar mass prior); (2) \citet{Faramaz+2021}; (3)  \citet{bainesetal2012}}
\tablenotetext{a}{Using an informed Gaussian stellar mass prior results in 1.46$^{+0.11}_{-0.15}$ \Msun.}
\tablenotetext{b}{\citet{Boyajian+2013} find a consistent value of $T_\mathrm{eff} = 7163 \pm 84$ K using the stellar angular diameter from \citet{bainesetal2012} together with their own calculation of the stellar bolometric flux.}
\end{deluxetable} 

\begin{figure*}[ht!]
  \includegraphics[trim=0cm 0cm 0cm 0cm, clip, width=7.in]{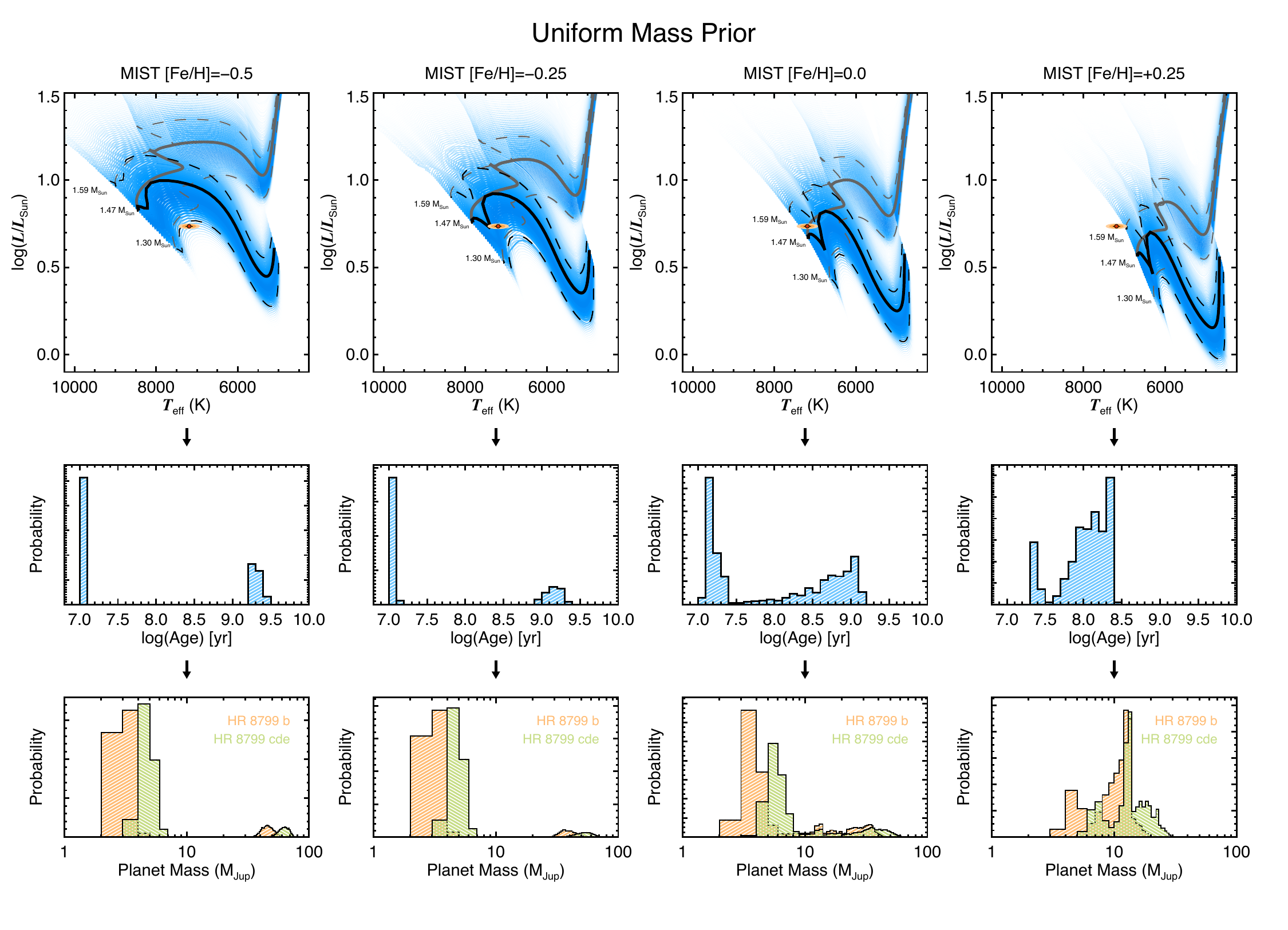}
  \centering
  \caption{\textit{Top Row}: MESA iso-mass tracks used to construct the age distributions of HR~8799. The thickness and color of each track is proportional to the probability density of the joint dynamical mass posterior distribution of HR~8799 at that model value. The solid black and gray line denotes the evolutionary track corresponding to the median value of our dynamical mass posterior. It is shown in black for the pre-main sequence phase of evolution and then becomes gray for the post-main sequence phase. Similarly, the dashed black and gray lines correspond to the 68\% HCI. Each column of panels corresponds to a different metallicity from [Fe/H] = --0.5 to +0.25 dex. The orange ellipse represents the measurement and 1-, 2-, and 3-$\sigma$ uncertainties of $L_{\star}$ and $T_{\rm eff}$ for HR~8799.
  \textit{Middle Row}: The inferred age distributions of HR~8799 following the method described in \S \ref{subsec:stellar}. \textit{Bottom Row}: Corresponding hot-start masses for the planets derived directly from the age distribution and planet luminosities. \label{fig:age_uniformprior}} 
\end{figure*}

We begin our assessment of the bulk metallicity and age by randomly drawing a stellar mass rounded to the nearest 0.01~\Msun\ with a probability proportional to the dynamical mass posterior distribution of HR 8799 found in \S\ref{sec:results}. We also draw a luminosity from a normal distribution with a mean of 5.441~\Lsun\ and a standard deviation of 0.066~\Lsun. The iso-mass track will either be inconsistent with the given luminosity throughout the entire evolution of that star, or it will intersect the given luminosity at least once.  If the former occurs, that trial is ignored and another mass is drawn; for the latter, ages are interpolated across the grid sampling at the luminosity crossing point.  If the corresponding model radius and temperature at that point are within 3$\sigma$ of the measured values of 1.502~$\pm$~0.035~\Rsun\ and 7193~$\pm$~87~K, respectively, then the ages are retained.  This process of drawing a mass and luminosity and saving ages consistent with the models is repeated 10$^4$ times to build up a distribution of ages consistent with the measured values. Results for the linearly uniform mass prior are shown in Figure~\ref{fig:age_uniformprior}. 

Our use of four different [Fe/H] values enables a qualitative constraint of the stellar bulk metallicity: the stellar parameters are most consistent with [Fe/H]$\sim$ --0.25--0.00 dex but only marginally consistent with [Fe/H]$=$ --0.5 and +0.25 dex. The $\lambda$ Bo\"{o}tis nature of HR~8799 has made previous attempts to characterize the bulk metallcity difficult because the surface abundances may not be an accurate tracer of the internal composition \citep[e.g.,][]{moya2010b,Murphy+2021}. In spite of the coarse sampling in the metallicities as well as potential systematic uncertainties in the MESA grids themselves, these qualitative constraints support a picture in which HR~8799 may have a solar-like or slightly sub-solar bulk composition.

There are two general solutions for the age of HR~8799 that correspond to an ambiguity related to whether it is moving toward or evolving from the zero age main sequence. For the [Fe/H]$=$ 0.00 dex grids, these correspond to ages of $\sim$10$^{7-7.4}$ yr and $\sim$10$^{8-9.1}$ yr. \citet{Fabrycky+Murray-Clay2010} found that the outer three HR~8799 planets must be $\lesssim$20 \Mjup\ to remain stable on timescales of tens of Myr. Motivated by this dynamical stability requirement, we ignore the older age scenario because this results in implausibly high planet masses up to the hydrogen-burning limit. For each metallicity grid, we discard ages corresponding to a planet mass cutoff of $>$20 \Mjup\ using hot-start evolutionary models (see \S\ref{sec:hotstartplanetmasses}).

Excluding the [Fe/H]$=$ +0.25 dex grids, which are least consistent with the measured properties of HR~8799, an age of 10--23 Myr best agrees with our dynamical mass. This is younger than the age of 38--48 Myr \citep{Bell+2015} that is associated with HR~8799 if it is a member of the Columba moving group. This is also somewhat younger than the hot-start cooling age of $42^{+24}_{-16}$ Myr for HR~8799~e recently derived by \citet{Brandt+2021DMe}, although at twice their quoted uncertainties, both ages are consistent. \citet{bainesetal2012} use Yonsei-Yale grids fixed at solar metallicity to derive an age consistent with either $\sim$33 Myr or $\sim$90 Myr. Regardless of kinematic association with a young moving group, these model-based constraints all support the picture that the HR~8799 system is $<$100 Myr and probably younger than 50 Myr.

To assess the impact of mass priors and restricted eccentricities on the results, we repeat this exercise for the other three scenarios considered in \S\ref{sec:results}: the dynamical mass with the informed Gaussian prior, and the two low-eccentricity cases. Summary statistics for the resulting trimmed distributions are tabulated in Table \ref{tab:hr8799_mist}. Figures showing the ages and corresponding hot-start masses using our other dynamical masses under different assumptions for the mass prior and range of eccentricities can be found in Appendix \ref{appendix:MESAGrids}. The values of the ages and hot-start masses do not significantly change depending on which dynamical mass is used, and the constraints of [Fe/H] also remain consistent.  

\begin{deluxetable}{cccc}
\renewcommand\arraystretch{0.9}
\tabletypesize{\small}
\setlength{ \tabcolsep } {.1cm} 
\tablewidth{0pt}
\tablecolumns{4}
\tablecaption{HR 8799 Age and Planet Masses from MIST Models\tablenotemark{a} \label{tab:hr8799_mist}}
\tablehead{
    \colhead{Model} & \colhead{} & \colhead{Age}  & \colhead{Mass} \\  
    \colhead{[Fe/H]} & \colhead{Planet} & \colhead{(Myr)} & \colhead{(M$_\mathrm{Jup}$)}   }   
\startdata
\cutinhead{Uniform Mass Prior}
+0.25 & b & 120$^{+130}_{-40}$ & 12$^{+2}_{-3}$ \\
0.0 & b & 17$^{+6}_{-5}$ & 4.0$^{+0.9}_{-1.3}$ \\
--0.25 & b & 11.3$^{+1.2}_{-1.2}$ & 3.1$^{+0.4}_{-0.4}$ \\
--0.50 & b & 11.3$^{+1}_{-1.2}$ & 3.1$^{+0.4}_{-0.4}$ \\
+0.25 & c/d/e & 120$^{+130}_{-40}$ & 13.2$^{+5.3}_{-1.5}$ \\
0.0 & c/d/e & 17$^{+6}_{-5}$ & 6.0$^{+1}_{-1.4}$ \\
--0.25 & c/d/e & 11.3$^{+1.2}_{-1.2}$ & 4.8$^{+0.5}_{-0.7}$ \\
--0.50 & c/d/e & 11.3$^{+1}_{-1.2}$ & 4.8$^{+0.5}_{-0.7}$ \\
\cutinhead{Normal Mass Prior}
+0.25 & b & 140$^{+110}_{-50}$ & 12.1$^{+1.5}_{-3.5}$ \\
0.0 & b & 17$^{+6}_{-5}$ & 4.0$^{+0.9}_{-1.2}$ \\
--0.25 & b & 11.3$^{+1.2}_{-1.2}$ & 3.1$^{+0.4}_{-0.4}$ \\
--0.50 & b & 11.2$^{+0.98}_{-1.2}$ & 3.0$^{+0.4}_{-0.4}$ \\
+0.25 & c/d/e & 140$^{+110}_{-50}$ & 13.4$^{+6.3}_{-0.84}$ \\
0.0 & c/d/e & 17$^{+6}_{-5}$ & 6.0$^{+1}_{-1.4}$ \\
--0.25 & c/d/e & 11.3$^{+1.2}_{-1.2}$ & 4.8$^{+0.5}_{-0.7}$ \\
--0.50 & c/d/e & 11.2$^{+0.98}_{-1.2}$ & 4.8$^{+0.5}_{-0.7}$ \\
\cutinhead{Uniform Mass Prior, Low Eccentricities}
+0.25 & b & 140$^{+100}_{-40}$ & 12.3$^{+1.3}_{-2.4}$ \\
0.0 & b & 19$^{+36}_{-9}$ & 4.3$^{+3.4}_{-1.9}$ \\
--0.25 & b & 11.3$^{+1.2}_{-1.2}$ & 3.1$^{+0.4}_{-0.4}$ \\
--0.50 & b & 11$^{+1}_{-1.1}$ & 3.1$^{+0.4}_{-0.4}$ \\
+0.25 & c/d/e & 140$^{+100}_{-40}$ & 13.5$^{+4.8}_{-1.1}$ \\
0.0 & c/d/e & 19$^{+36}_{-9}$ & 6.2$^{+1.1}_{-1.6}$ \\
--0.25 & c/d/e & 11.3$^{+1.2}_{-1.2}$ & 4.8$^{+0.5}_{-0.7}$ \\
--0.50 & c/d/e & 11.2$^{+1}_{-1.1}$ & 4.8$^{+0.5}_{-0.7}$ \\
\cutinhead{Normal Mass Prior, Low Eccentricities}
+0.25 & b & 120$^{+150}_{-57}$ & 12$^{+2}_{-5}$ \\
0.0 & b & 19$^{+4}_{-7}$ & 4.3$^{+2.9}_{-1.9}$ \\
--0.25 & b & 11.3$^{+1.2}_{-1.2}$ & 3.1$^{+0.4}_{-0.4}$ \\
--0.50 & b & 11.2$^{+1.0}_{-1.2}$ & 3.0$^{+0.4}_{-0.4}$ \\
+0.25 & c/d/e & 120$^{+150}_{-57}$ & 13.1$^{+1.9}_{-6.7}$ \\
0.0 & c/d/e & 19$^{+4}_{-7}$ & 6.2$^{+1.1}_{-1.5}$ \\
--0.25 & c/d/e & 11.3$^{+1.2}_{-1.2}$ & 4.8$^{+0.5}_{-0.7}$ \\
--0.50 & c/d/e & 11.2$^{+1.0}_{-1.2}$ & 4.8$^{+0.5}_{-0.7}$ \\
\enddata
\tablenotetext{a}{Assumes dynamical stability in which planet masses $>$20~M$_\mathrm{Jup}$ are excluded.}
\end{deluxetable}

\subsection{Hot-Start Mass Constraints of the Planets}\label{sec:hotstartplanetmasses}
In light of these revised age estimates, we use hot-start evolutionary models from \citet{Burrows+1997} to infer the masses of the imaged companions using their  bolometric luminosities (Figure \ref{fig:age_uniformprior}). Here we assume the age of the star to be the age of the planets. For planet b, we adopt a bolometric luminosity of log$(L_{\rm bol}/{\Lsun})$ = --5.1 $\pm$ 0.1 dex, and for planets c and d we adopt log$(L_{\rm bol}/{\Lsun})$ = --4.7 $\pm$ 0.1 dex  \citep{marios2008,Rajan+2015}. The spectrum and near-infrared contrast of planet e is similar to that of planets c and d \citep[e.g.,][]{marios2010,greenbaumetal2018}, thus we adopt the same value of log$(L_{\rm bol}/{\Lsun})$ = --4.7 $\pm$ 0.1 dex for this inner planet.

Excluding the [Fe/H]$=$ +0.25 dex grids and the discarded solutions corresponding to old ($\sim$Gyr) ages, the resulting mass estimates are 2.7--4.9 \Mjup\ for HR~8799~b and 4.1--7.0 \Mjup\ for HR~8799~c, d, and e (Table \ref{tab:hr8799_mist}). These masses are within the giant planet regime assuming a boundary of $\approx$13 \Mjup\ and are lower than, but still broadly consistent with, many of the previous estimates in the literature. For example, \citet{marios2008} and \citet{marios2010} report hot-start masses of 5--7 \Mjup\ for planet b and 7--10 \Mjup\ for planets c, d, and e for ages of 30 Myr and 60 Myr. \citet{wilner2018} estimated a dynamical mass of HR~8799~b to be 5.8$^{+7.9}_{-3.1}$ \Mjup\ if it is responsible for clearing mass in its chaotic zone out to the inner edge of the debris disk. Within their subset of dynamically stable orbital solutions, \citet{Wang+2018} find hot-start masses of 5.8$\pm$0.5 \Mjup\ for planet b and 7.2$^{+0.6}_{-0.7}$ \Mjup\ for planets c, d, and e for an age of 42$\pm$5 Myr. \citet{Brandt+2021DMe} recently presented a dynamical mass of 9.6$^{+1.9}_{-1.8}$ \Mjup\ for HR~8799~e by using a \textit{Gaia} EDR3 host star proper motion anomaly together with results from \citet{Wang+2018} for orbital solutions and planet mass ratios. This is somewhat higher than the values we found and could be caused by an underestimated age, compositional difference between the star and planet, higher planet luminosity, or systematic errors in the models themselves. Altogether, the mass estimates presented in this work imply a minimum-mass extrasolar nebula of $M\geq$15--26 \Mjup\ for HR~8799.

\section{Conclusion} 
We have measured the model-independent dynamical mass of HR 8799 in a Bayesian framework using all available astrometry of its four planets and treating them as massless independent particles. The joint dynamical mass of HR~8799 is 1.47$^{+0.12}_{-0.17}$ \Msun\ assuming a uniform prior, validating regularly used values of $\approx$1.5 \Msun\ in previous studies. The modest but realistic uncertainties improve to 1.46$^{+0.11}_{-0.15}$ \Msun\ when an informative prior based on the stellar spectroscopy is used. When only near-circular orbits of $e<$0.1 are allowed, the joint dynamical mass measurement becomes 1.43$^{+0.06}_{-0.07}$ \Msun.

We used our stellar mass constraints together with other previously measured parameters of the host star to investigate the age and bulk metallicity of HR 8799. The age constraint from MESA models is 10--23 Myr after excluding ages corresponding to hot-start masses of $\geq$20 \Mjup. This supports an intermediate age for this system independent of kinematic membership status to young moving groups. We also find that the favored bulk metallicity is [Fe/H]$\sim$ --0.25--0.00 dex.

Finally, using these inferred ages, we derived hot start masses of 2.7--4.9 \Mjup\ for HR 8799 b and 4.1--7.0 \Mjup\ for HR 8799 c, d, and e. These are somewhat lower than typical estimates, as expected from the younger inferred age. Continued astrometric monitoring of this system will enable future studies to progressively reduce the uncertainty in the orbital elements of the planets and the mass of HR~8799. However, eventually the assumption that $M_{\star} \approx M_{tot}$ will break down as the total mass precision becomes comparable to the planet masses. At that point, a more complete dynamical modeling that includes mutual planet interactions will be needed to properly account for the total system mass interior to each orbit \citep[e.g.,][]{Lacour+2021}.   

\acknowledgments
\textit{Acknowledgments}. We thank the anonymous referee for constructive suggestions as well as Henry Ngo and Sarah Blunt for their consulting with the \texttt{orbitize!} package. 

A.G.S acknowledges support from the Texas Astronomy Undergraduate Research experience for Under-represented Students (TAURUS) Scholars program. TAURUS graciously acknowledges funding from the National Science Foundation (NSF), National Aeronautics and Space Administration (NASA), the Cox Board of Visitors for the University of Texas’ Department of Astronomy and McDonald Observatory, and many generous donors who gave to the TAURUS HornRaiser campaign. TAURUS is also grateful for the countless time commitments of students, staff, and faculty at UT to build a healthier, happier, and more representative astronomy community. This material is based upon work supported by the National Science Foundation Graduate Research Fellowship Program under Grant No. 1842402. B.P.B. acknowledges support from the National Science Foundation grant AST-1909209 and NASA Exoplanet Research Program grant 20-XRP20 2-0119. 

This work received computational support from UTSA's HPC cluster SHAMU, operated by the Office of Information Technology. A.G.S. thanks Eric Schlegel for the sponsorship that enabled the use of this resource. 

Some of the data presented herein were obtained at the W. M. Keck Observatory, which is operated as a scientific partnership among the California Institute of Technology, the University of California and the National Aeronautics and Space Administration. The Observatory was made possible by the generous financial support of the W. M. Keck Foundation. The authors wish to recognize and acknowledge the very significant cultural role and reverence that the summit of Maunakea has always had within the indigenous Hawaiian community.  We are most fortunate to have the opportunity to conduct observations from this mountain. 

This work has made use of data from the European Space Agency (ESA) mission
{\it Gaia} (\url{https://www.cosmos.esa.int/gaia}), processed by the {\it Gaia}
Data Processing and Analysis Consortium (DPAC,
\url{https://www.cosmos.esa.int/web/gaia/dpac/consortium}). Funding for the DPAC
has been provided by national institutions, in particular the institutions
participating in the {\it Gaia} Multilateral Agreement. 

This research has made use of NASA's Astrophysics Data System Bibliographic Services.

\vspace{5mm}

\facility{Keck:II (NIRC2)}
\software{\texttt{orbitize!} \citep{orbitize}, \texttt{ptemcee} \citep{foremanmackey13,ptemcee}, \texttt{matplotlib} \citep{matplotlib}, \texttt{numpy} \citep{numpy,Harris+2020}, \texttt{astropy} \citep{astropy,astropy2018}}

\appendix

\section{Orbit Fitting Results}\label{appendix}

\begin{figure*}
  \includegraphics[trim=0cm 0cm 8.6cm 0cm, clip, width=7.1in]{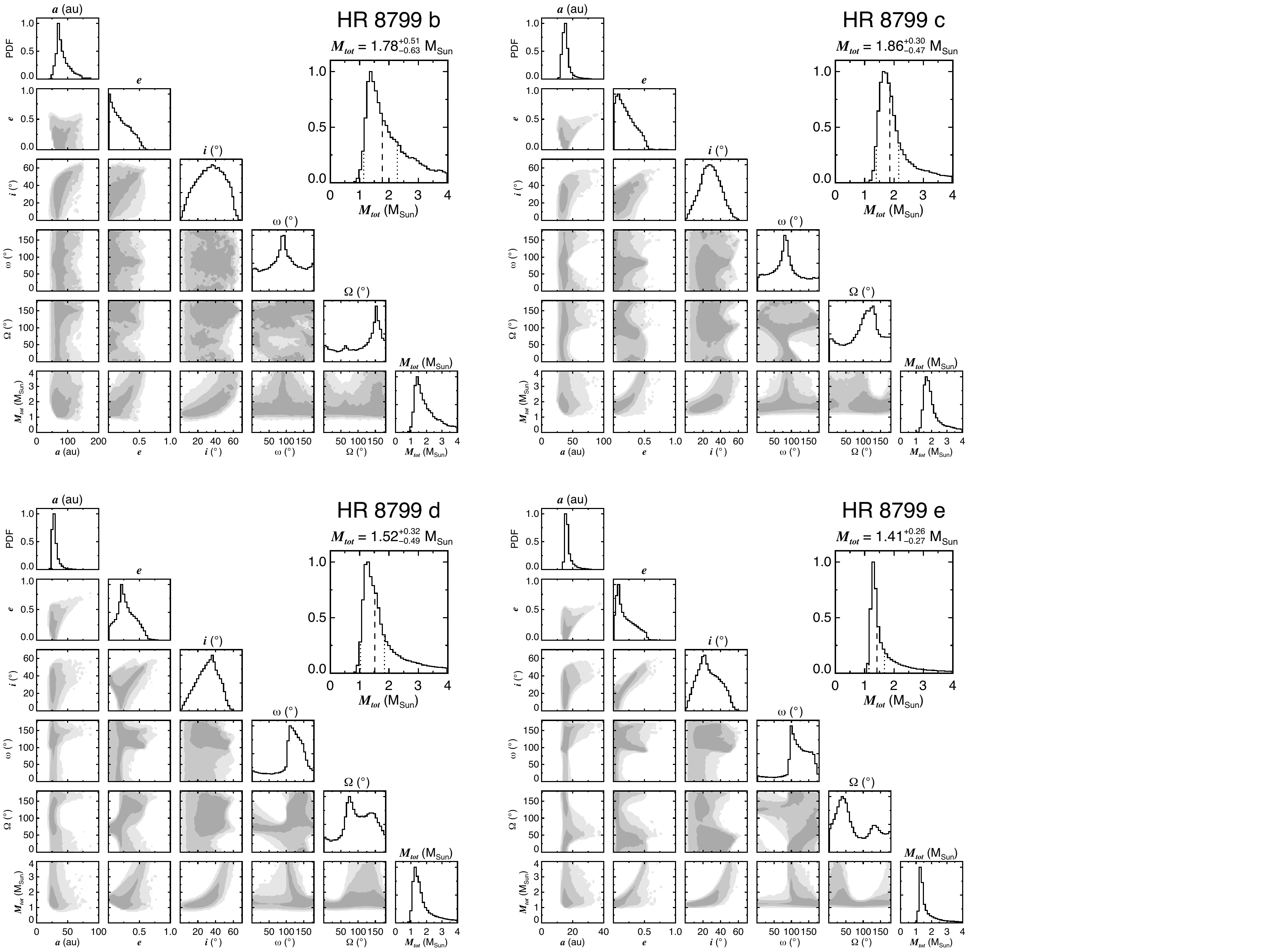}
  \centering
  \caption{Summary of the MCMC orbit-fitting results for a uniform prior on stellar host mass. \label{fig:quadcorners}} 
\end{figure*}

 Corner plots showing the two-dimensional joint posterior distributions for each combination of parameters from our MCMC orbit fit of HR~8799 b, c, d, and e are shown in Figure \ref{fig:quadcorners}. These results are for a uniform prior on stellar mass and summarized in Table \ref{tab:hr8799_orbit}.  The marginalized posterior distributions for the orbital parameters are displayed along the diagonal, and $M_{tot}$ is highlighted in the upper right. The contours of the 2-D histograms are 68.3\%, 95.4\%, and 99.7\% density levels.

\begin{deluxetable*}{lcccccc}
\renewcommand\arraystretch{0.9}
\tabletypesize{\small}
\setlength{ \tabcolsep } {.1cm} 
\tablewidth{0pt}
\tablecolumns{5}
\tablecaption{HR 8799 Orbit Fit Results \label{tab:hr8799_orbit}}
\tablehead{
    \colhead{Parameter} & \colhead{Median}  & \colhead{MAP\tablenotemark{a}}  & \colhead{68.3\% CI}  & \colhead{95.4\% CI}  
        }   
\startdata
\cutinhead{HR 8799 b}
$a$ (AU) & 77.9 & 70.5 & (59.6, 94.4) & (50.1, 132) \\
$e$ & 0.17 & 0.015 & (0.0, 0.27) & (0.0, 0.48) \\
$i$ ($^{\circ}$) & 35 & 37 & (20, 53) & (7, 60) \\
$\omega$ ($^{\circ}$) & 91 & 91 & (47, 155) & (10, 180) \\
$\Omega$ ($^{\circ}$) & 128 & 154 & (84, 180) & (9.6, 180) \\
$M_\mathrm{tot}$ (M$_{\odot}$) & 1.78 & 1.35 & (1.14, 2.29) & (1.0, 3.60) \\
\cutinhead{HR 8799 c}
$a$ (AU) & 38.2 & 39.5 & (32.3, 41.4) & (29.2, 55.3) \\
$e$ & 0.18 & 0.06 & (0.0, 0.27) & (0.0, 0.48) \\
$i$ ($^{\circ}$) & 28 & 25 & (17, 40) & (6, 51) \\
$\omega$ ($^{\circ}$) & 83 & 84 & (22, 116) & (2, 169) \\
$\Omega$ ($^{\circ}$) & 108 & 127 & (78, 173) & (10, 180) \\
$M_\mathrm{tot}$ (M$_{\odot}$) & 1.86 & 1.67 & (1.39, 2.16) & (1.26, 3.46) \\
\cutinhead{HR 8799 d}
$a$ (AU) & 28.5 & 26.5 & (24.4, 31.1) & (22.8, 42.9) \\
$e$ & 0.27 & 0.21 & (0.13, 0.44) & (0.0, 0.56) \\
$i$ ($^{\circ}$) & 32 & 36 & (20, 45) & (7, 52) \\
$\omega$ ($^{\circ}$) & 120 & 111 & (102, 161) & (17, 180) \\
$\Omega$ ($^{\circ}$) & 100 & 74 & (63, 151) & (18, 180) \\
$M_\mathrm{tot}$ (M$_{\odot}$) & 1.52 & 1.25 & (1.03, 1.84) & (0.93, 3.35) \\
\cutinhead{HR 8799 e}
$a$ (AU) & 16.2 & 15.5 & (14.5, 17.3) & (14.0, 23.9) \\
$e$ & 0.16 & 0.06 & (0.0, 0.26) & (0.0, 0.49) \\
$i$ ($^{\circ}$) & 25 & 22 & (12, 39) & (5, 51) \\
$\omega$ ($^{\circ}$) & 119 & 97 & (94, 154) & (30, 180) \\
$\Omega$ ($^{\circ}$) & 59 & 39 & (0.0, 105) & (0.074, 169) \\
$M_\mathrm{tot}$ (M$_{\odot}$) & 1.41 & 1.29 & (1.15, 1.68) & (1.09, 3.15) \\
\enddata
\tablenotetext{a}{Maximum a posteriori probability.}
\end{deluxetable*}

\section{Low-Eccentricity Orbit Fitting Results} \label{appendix:orbitFitLE}

\begin{figure*}[b!]
  \includegraphics[trim=0cm 0cm 8.6cm 0cm, clip, width=7.1in]{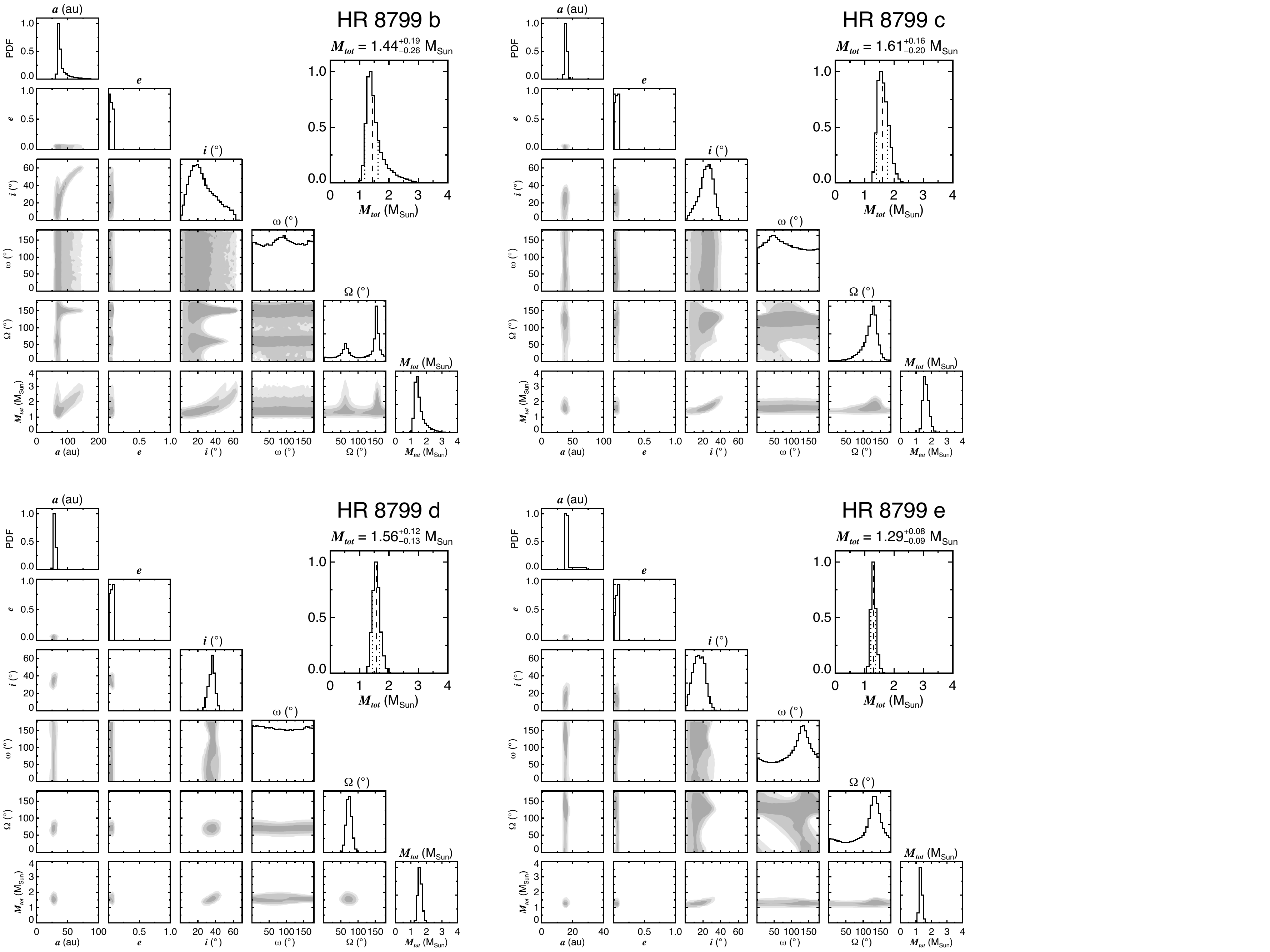}
  \centering
  \caption{Same as Figure \ref{fig:quadcorners} but for the low eccentricity scenario as described in \S \ref{sec:lowe}. \label{fig:quadcornersLE}} 
\end{figure*}

\begin{figure*}[ht!]
  \includegraphics[trim=0cm 9.5cm 0cm 0cm, clip, width=7.1in]{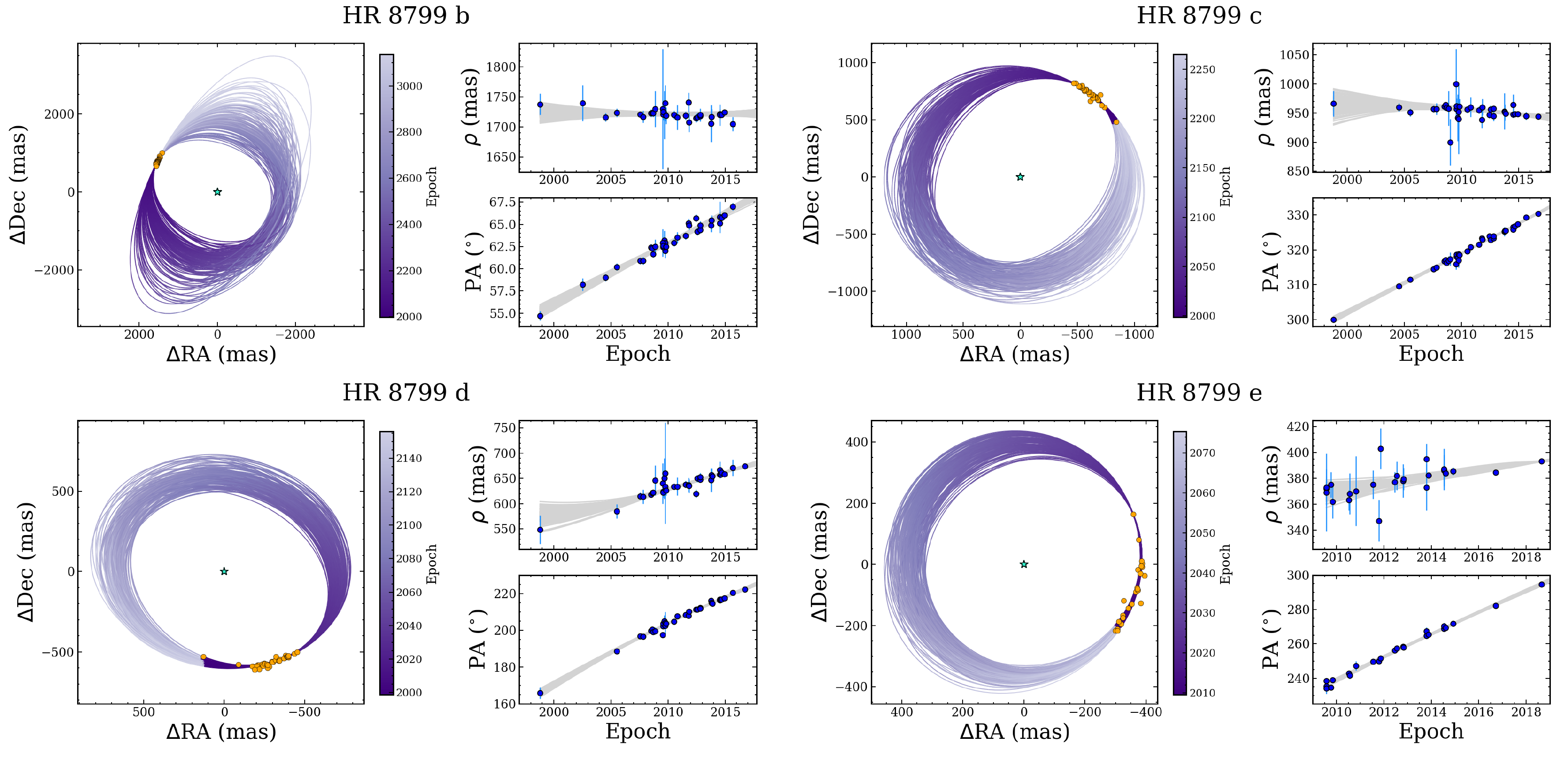}
  \centering
  \caption{Same as Figure \ref{fig:quadorbitsamples} but for the low-eccentricity scenario as described in \S\ref{sec:lowe}. \label{fig:quadorbitsamples_LE}} 
\end{figure*}

The results of our MCMC orbit fits for the low-eccentricity scenario are summarized in Figure \ref{fig:quadcornersLE} and Table \ref{tab:hr8799_orbit_le}. We also display a sample of the sky-projected orbits drawn from the MCMC chains in Figure \ref{fig:quadorbitsamples_LE}. 

\begin{deluxetable*}{lcccccc}
\renewcommand\arraystretch{0.9}
\tabletypesize{\small}
\setlength{ \tabcolsep } {.1cm} 
\tablewidth{0pt}
\tablecolumns{5}
\tablecaption{HR 8799 Orbit Fit Results (Low-Eccentricity Scenario) \label{tab:hr8799_orbit_le}}
\tablehead{
    \colhead{Parameter} & \colhead{Median}  & \colhead{MAP\tablenotemark{a}}  & \colhead{68.3\% CI}  & \colhead{95.4\% CI}  
        }   
\startdata
\cutinhead{HR 8799 b}
$a$ (AU) & 73.8 & 70.5 & (65.6, 79.5) & (64.1, 121) \\
$e$ & 0.044 & 0.0 & (0.0, 0.063) & (0.0, 0.094) \\
$i$ ($^{\circ}$) & 23 & 19 & (6.4, 34) & (3, 56) \\
$\omega$ ($^{\circ}$) & 90 & 95 & (60, 180) & (7.4, 179) \\
$\Omega$ ($^{\circ}$) & 136 & 154 & (62, 161) & (19, 180) \\
$M_\mathrm{tot}$ (M$_{\odot}$) & 1.44 & 1.35 & (1.18, 1.63) & (1.1, 2.31) \\
\cutinhead{HR 8799 c}
$a$ (AU) & 39.3 & 39.5 & (37.4, 40.9) & (36.1, 42.7) \\
$e$ & 0.053 & 0.06 & (0.034, 0.10) & (0.0, 0.10) \\
$i$ ($^{\circ}$) & 24 & 26 & (18, 33) & (7, 37) \\
$\omega$ ($^{\circ}$) & 83 & 50 & (1.5, 118) & (1, 172) \\
$\Omega$ ($^{\circ}$) & 124 & 131 & (110, 145) & (52, 167) \\
$M_\mathrm{tot}$ (M$_{\odot}$) & 1.61 & 1.51 & (1.41, 1.77) & (1.31, 2.01) \\
\cutinhead{HR 8799 d}
$a$ (AU) & 29.2 & 29.5 & (27.8, 30.5) & (26.6, 32.1) \\
$e$ & 0.053 & 0.095 & (0.034, 0.10) & (0.0, 0.10) \\
$i$ ($^{\circ}$) & 36 & 36 & (32, 39) & (30, 42) \\
$\omega$ ($^{\circ}$) & 88.9 & 17.5 & (0.00841, 123) & (0.036, 172) \\
$\Omega$ ($^{\circ}$) & 73.1 & 73 & (65, 80.9) & (58, 88.9) \\
$M_\mathrm{tot}$ (M$_{\odot}$) & 1.56 & 1.57 & (1.43, 1.68) & (1.3, 1.82) \\
\cutinhead{HR 8799 e}
$a$ (AU) & 16.0 & 15.5 & (15.4, 16.5) & (14.9, 17.1) \\
$e$ & 0.063 & 0.06 & (0.048, 0.10) & (0.0, 0.10) \\
$i$ ($^{\circ}$) & 16 & 14 & (9.1, 23) & (3, 27) \\
$\omega$ ($^{\circ}$) & 115 & 130 & (82, 180) & (10, 180) \\
$\Omega$ ($^{\circ}$) & 120 & 130 & (96, 174) & (13, 180) \\
$M_\mathrm{tot}$ (M$_{\odot}$) & 1.29 & 1.29 & (1.21, 1.37) & (1.14, 1.48) \\
\enddata
\tablenotetext{a}{Maximum a posteriori probability.}
\end{deluxetable*}

\newpage
\section{MESA Ages and Hot-Start Planet Masses}\label{appendix:MESAGrids}
\begin{figure*}[t!]
  \includegraphics[trim=0cm 0cm 0cm 0cm, clip, width=5.65in]{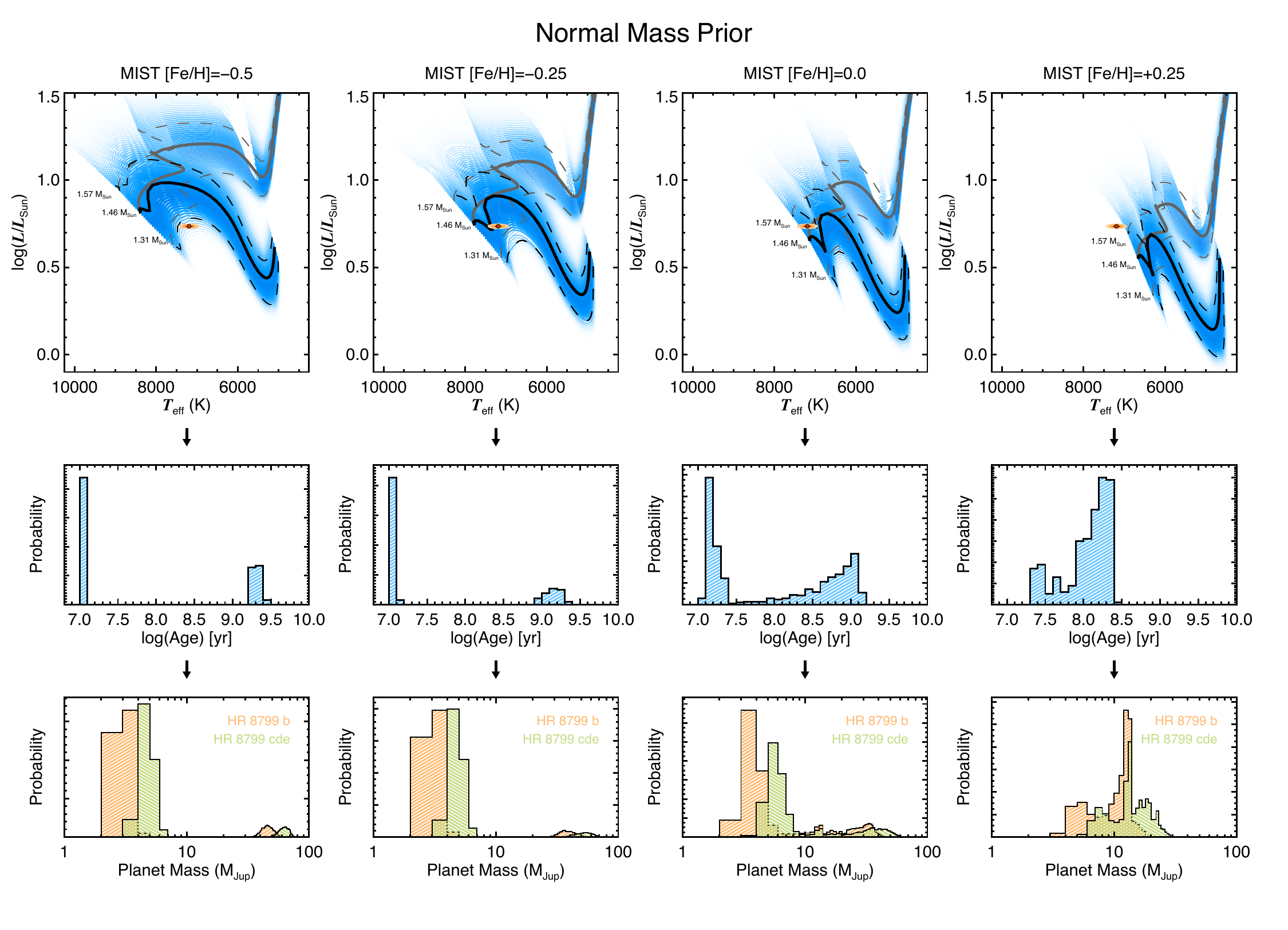}
  \centering
  \caption{Same as Figure \ref{fig:age_uniformprior} but using the dynamical mass with an informed Gaussian prior of 1.47 $\pm$ 0.30 \Msun\ for the stellar mass. \label{fig:age_gaussprior}} 
\end{figure*}

\begin{figure*}[h!]
  \includegraphics[trim=0cm 0cm 0cm 0cm, clip, width=5.65in]{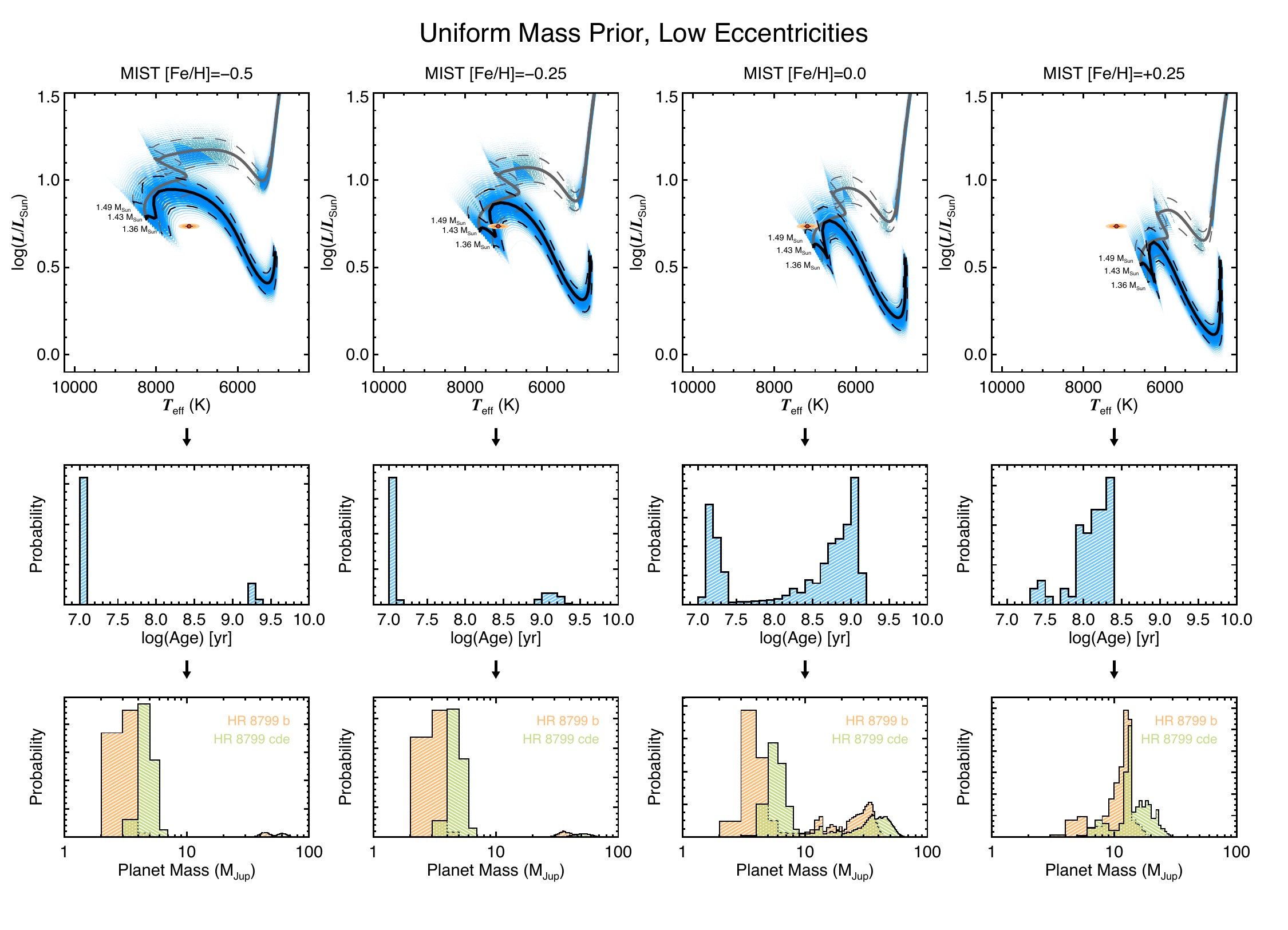}
  \centering
  \caption{Same as Figure \ref{fig:age_uniformprior} but using the dynamical mass with a uniform prior and low eccentricities. \label{fig:age_uniformprior_lowecc}} 
\end{figure*}

\begin{figure*}[h!]
  \includegraphics[trim=0cm 0cm 0cm 0cm, clip, width=5.65in]{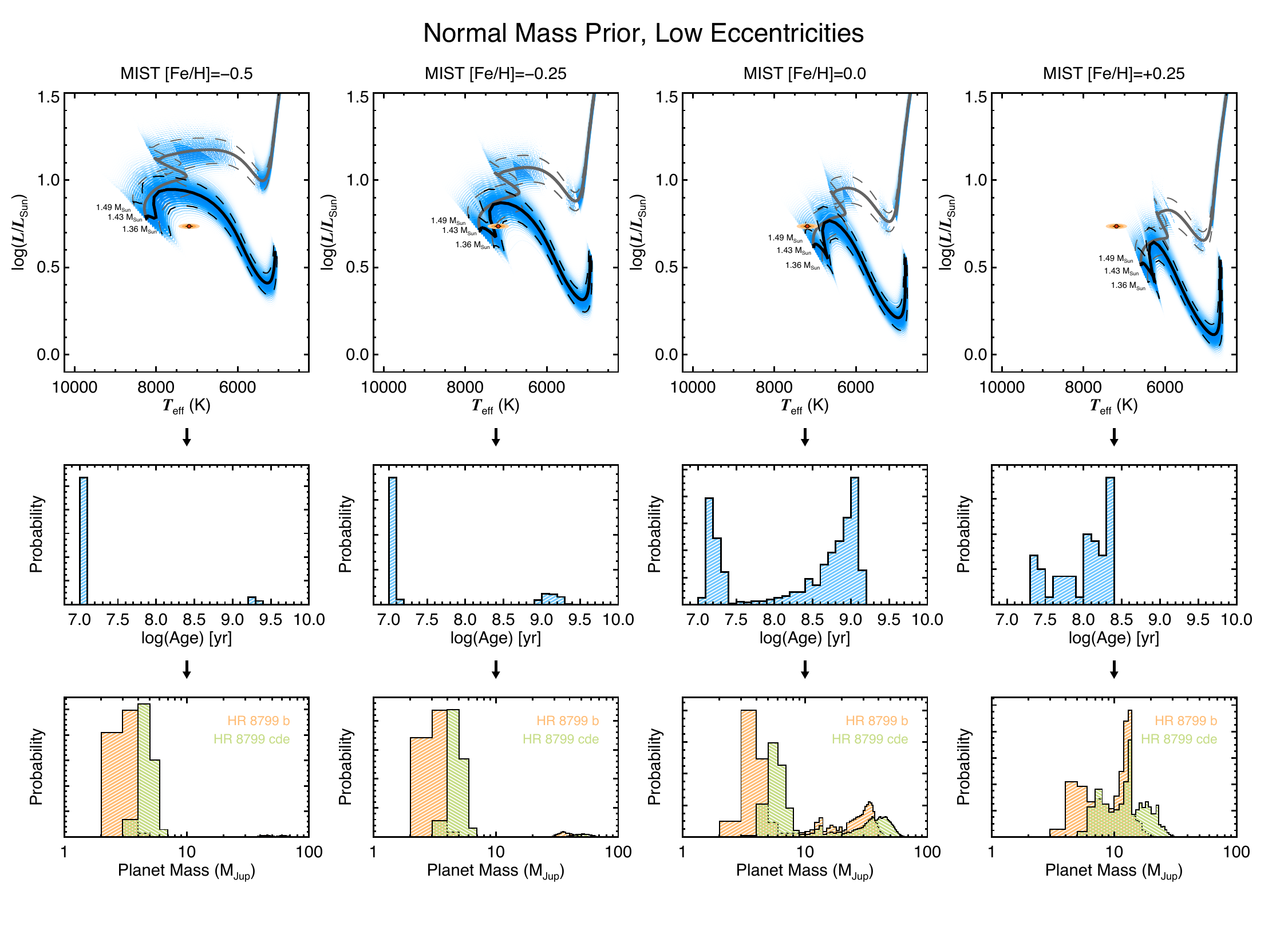}
  \centering
  \caption{Same as Figure \ref{fig:age_uniformprior} but using the dynamical mass with an informed Gaussian prior of 1.47 $\pm$ 0.30 \Msun\ for the stellar mass and low eccentricities.  \label{fig:age_gaussprior_lowecc}} 
\end{figure*}

This Appendix includes a summary of results from stellar and substellar evolutionary models following our procedure described in \S\ref{subsec:stellar} but using the other three dynamical masses from \S\ref{sec:results}. Figure \ref{fig:age_gaussprior} shows results using the dynamical mass with the informed Gaussian prior. Results with the low-eccentricity assumption using the dynamical mass with a uniform prior and with the informed Gaussian prior are displayed in Figures \ref{fig:age_uniformprior_lowecc} and \ref{fig:age_gaussprior_lowecc}, respectively.

\clearpage

\bibliographystyle{aasjournal}
\bibliography{ms}

\begin{thebibliography}{}
\expandafter\ifx\csname natexlab\endcsname\relax\def\natexlab#1{#1}\fi
\providecommand{\url}[1]{\href{#1}{#1}}
\providecommand{\dodoi}[1]{doi:~\href{http://doi.org/#1}{\nolinkurl{#1}}}
\providecommand{\doeprint}[1]{\href{http://ascl.net/#1}{\nolinkurl{http://ascl.net/#1}}}
\providecommand{\doarXiv}[1]{\href{https://arxiv.org/abs/#1}{\nolinkurl{https://arxiv.org/abs/#1}}}

\bibitem[{{Apai} {et~al.}(2016){Apai}, {Kasper}, {Skemer}, {Hanson},
  {Lagrange}, {Biller}, {Bonnefoy}, {Buenzli}, \& {Vigan}}]{Apai+2016}
{Apai}, D., {Kasper}, M., {Skemer}, A., {et~al.} 2016, \apj, 820, 40,
  \dodoi{10.3847/0004-637X/820/1/40}

\bibitem[{{Astropy Collaboration} {et~al.}(2013){Astropy Collaboration},
  {Robitaille}, {Tollerud}, {Greenfield}, {Droettboom}, {Bray}, {Aldcroft},
  {Davis}, {Ginsburg}, {Price-Whelan}, {Kerzendorf}, {Conley}, {Crighton},
  {Barbary}, {Muna}, {Ferguson}, {Grollier}, {Parikh}, {Nair}, {Unther},
  {Deil}, {Woillez}, {Conseil}, {Kramer}, {Turner}, {Singer}, {Fox}, {Weaver},
  {Zabalza}, {Edwards}, {Azalee Bostroem}, {Burke}, {Casey}, {Crawford},
  {Dencheva}, {Ely}, {Jenness}, {Labrie}, {Lim}, {Pierfederici}, {Pontzen},
  {Ptak}, {Refsdal}, {Servillat}, \& {Streicher}}]{astropy}
{Astropy Collaboration}, {Robitaille}, T.~P., {Tollerud}, E.~J., {et~al.} 2013,
  \aap, 558, A33, \dodoi{10.1051/0004-6361/201322068}

\bibitem[{{Astropy Collaboration} {et~al.}(2018){Astropy Collaboration},
  {Price-Whelan}, {Sip{\H o}cz}, {G{\"u}nther}, {Lim}, {Crawford}, {Conseil},
  {Shupe}, {Craig}, {Dencheva}, {Ginsburg}, {VanderPlas}, {Bradley},
  {P{\'e}rez-Su{\'a}rez}, {de Val-Borro}, {Aldcroft}, {Cruz}, {Robitaille},
  {Tollerud}, {Ardelean}, {Babej}, {Bach}, {Bachetti}, {Bakanov}, {Bamford},
  {Barentsen}, {Barmby}, {Baumbach}, {Berry}, {Biscani}, {Boquien}, {Bostroem},
  {Bouma}, {Brammer}, {Bray}, {Breytenbach}, {Buddelmeijer}, {Burke},
  {Calderone}, {Cano Rodr{\'{\i}}guez}, {Cara}, {Cardoso}, {Cheedella},
  {Copin}, {Corrales}, {Crichton}, {D'Avella}, {Deil}, {Depagne}, {Dietrich},
  {Donath}, {Droettboom}, {Earl}, {Erben}, {Fabbro}, {Ferreira}, {Finethy},
  {Fox}, {Garrison}, {Gibbons}, {Goldstein}, {Gommers}, {Greco}, {Greenfield},
  {Groener}, {Grollier}, {Hagen}, {Hirst}, {Homeier}, {Horton}, {Hosseinzadeh},
  {Hu}, {Hunkeler}, {Ivezi{\'c}}, {Jain}, {Jenness}, {Kanarek}, {Kendrew},
  {Kern}, {Kerzendorf}, {Khvalko}, {King}, {Kirkby}, {Kulkarni}, {Kumar},
  {Lee}, {Lenz}, {Littlefair}, {Ma}, {Macleod}, {Mastropietro}, {McCully},
  {Montagnac}, {Morris}, {Mueller}, {Mumford}, {Muna}, {Murphy}, {Nelson},
  {Nguyen}, {Ninan}, {N{\"o}the}, {Ogaz}, {Oh}, {Parejko}, {Parley}, {Pascual},
  {Patil}, {Patil}, {Plunkett}, {Prochaska}, {Rastogi}, {Reddy Janga},
  {Sabater}, {Sakurikar}, {Seifert}, {Sherbert}, {Sherwood-Taylor}, {Shih},
  {Sick}, {Silbiger}, {Singanamalla}, {Singer}, {Sladen}, {Sooley},
  {Sornarajah}, {Streicher}, {Teuben}, {Thomas}, {Tremblay}, {Turner},
  {Terr{\'o}n}, {van Kerkwijk}, {de la Vega}, {Watkins}, {Weaver}, {Whitmore},
  {Woillez}, {Zabalza}, \& {Astropy Contributors}}]{astropy2018}
{Astropy Collaboration}, {Price-Whelan}, A.~M., {Sip{\H o}cz}, B.~M., {et~al.}
  2018, \aj, 156, 123, \dodoi{10.3847/1538-3881/aabc4f}

\bibitem[{{Baines} {et~al.}(2012){Baines}, {White}, {Huber}, {Jones},
  {Boyajian}, {McAlister}, {ten Brummelaar}, {Turner}, {Sturmann}, {Sturmann},
  {Goldfinger}, {Farrington}, {Riedel}, {Ireland}, {von Braun}, \&
  {Ridgway}}]{bainesetal2012}
{Baines}, E.~K., {White}, R.~J., {Huber}, D., {et~al.} 2012, \apj, 761, 57,
  \dodoi{10.1088/0004-637X/761/1/57}

\bibitem[{{Bell} {et~al.}(2015){Bell}, {Mamajek}, \& {Naylor}}]{Bell+2015}
{Bell}, C. P.~M., {Mamajek}, E.~E., \& {Naylor}, T. 2015, \mnras, 454, 593,
  \dodoi{10.1093/mnras/stv1981}

\bibitem[{{Biller} {et~al.}(2021){Biller}, {Apai}, {Bonnefoy}, {Desidera},
  {Gratton}, {Kasper}, {Kenworthy}, {Lagrange}, {Lazzoni}, {Mesa}, {Vigan},
  {Wagner}, {Vos}, \& {Zurlo}}]{Biller+2021}
{Biller}, B.~A., {Apai}, D., {Bonnefoy}, M., {et~al.} 2021, \mnras, 503, 743,
  \dodoi{10.1093/mnras/stab202}

\bibitem[{{Blunt} {et~al.}(2020){Blunt}, {Wang}, {Angelo}, {Ngo}, {Cody}, {De
  Rosa}, {Graham}, {Hirsch}, {Nagpal}, {Nielsen}, {Pearce}, {Rice}, \&
  {Tejada}}]{orbitize}
{Blunt}, S., {Wang}, J.~J., {Angelo}, I., {et~al.} 2020, \aj, 159, 89,
  \dodoi{10.3847/1538-3881/ab6663}

\bibitem[{{Bohn} {et~al.}(2020{\natexlab{a}}){Bohn}, {Kenworthy}, {Ginski},
  {Manara}, {Pecaut}, {de Boer}, {Keller}, {Mamajek}, {Meshkat}, {Reggiani},
  {Todorov}, \& {Snik}}]{Bohn+2020a}
{Bohn}, A.~J., {Kenworthy}, M.~A., {Ginski}, C., {et~al.} 2020{\natexlab{a}},
  \mnras, 492, 431, \dodoi{10.1093/mnras/stz3462}

\bibitem[{{Bohn} {et~al.}(2020{\natexlab{b}}){Bohn}, {Kenworthy}, {Ginski},
  {Rieder}, {Mamajek}, {Meshkat}, {Pecaut}, {Reggiani}, {de Boer}, {Keller},
  {Snik}, \& {Southworth}}]{Bohn+2020b}
---. 2020{\natexlab{b}}, \apjl, 898, L16, \dodoi{10.3847/2041-8213/aba27e}

\bibitem[{{Booth} {et~al.}(2016){Booth}, {Jord{\'a}n}, {Casassus}, {Hales},
  {Dent}, {Faramaz}, {Matr{\`a}}, {Barkats}, {Brahm}, \&
  {Cuadra}}]{boothetal2016}
{Booth}, M., {Jord{\'a}n}, A., {Casassus}, S., {et~al.} 2016, \mnras, 460, L10,
  \dodoi{10.1093/mnrasl/slw040}

\bibitem[{{Bowler}(2016)}]{bowler2016}
{Bowler}, B.~P. 2016, \pasp, 128, 102001,
  \dodoi{10.1088/1538-3873/128/968/102001}

\bibitem[{{Bowler} {et~al.}(2020){Bowler}, {Blunt}, \& {Nielsen}}]{Bowler+2020}
{Bowler}, B.~P., {Blunt}, S.~C., \& {Nielsen}, E.~L. 2020, \aj, 159, 63,
  \dodoi{10.3847/1538-3881/ab5b11}

\bibitem[{Bowler {et~al.}(2015)Bowler, Liu, Shkolnik, \&
  Tamura}]{Bowler:2015ja}
Bowler, B.~P., Liu, M.~C., Shkolnik, E.~L., \& Tamura, M. 2015, ApJS, 216, 7

\bibitem[{Bowler {et~al.}(2018)Bowler, Dupuy, Endl, Cochran, MacQueen, Fulton,
  Petigura, Howard, Hirsch, Kratter, Crepp, Biller, Johnson, \&
  Wittenmyer}]{Bowler:2018gy}
Bowler, B.~P., Dupuy, T.~J., Endl, M., {et~al.} 2018, AJ, 155, 159

\bibitem[{{Boyajian} {et~al.}(2013){Boyajian}, {von Braun}, {van Belle},
  {Farrington}, {Schaefer}, {Jones}, {White}, {McAlister}, {ten Brummelaar},
  {Ridgway}, {Gies}, {Sturmann}, {Sturmann}, {Turner}, {Goldfinger}, \&
  {Vargas}}]{Boyajian+2013}
{Boyajian}, T.~S., {von Braun}, K., {van Belle}, G., {et~al.} 2013, \apj, 771,
  40, \dodoi{10.1088/0004-637X/771/1/40}

\bibitem[{{Brandt} {et~al.}(2021{\natexlab{a}}){Brandt}, {Brandt}, {Dupuy},
  {Li}, \& {Michalik}}]{Brandt+2021}
{Brandt}, G.~M., {Brandt}, T.~D., {Dupuy}, T.~J., {Li}, Y., \& {Michalik}, D.
  2021{\natexlab{a}}, \aj, 161, 179, \dodoi{10.3847/1538-3881/abdc2e}

\bibitem[{{Brandt} {et~al.}(2021{\natexlab{b}}){Brandt}, {Brandt}, {Dupuy},
  {Michalik}, \& {Marleau}}]{Brandt+2021DMe}
{Brandt}, G.~M., {Brandt}, T.~D., {Dupuy}, T.~J., {Michalik}, D., \& {Marleau},
  G.-D. 2021{\natexlab{b}}, \apjl, 915, L16, \dodoi{10.3847/2041-8213/ac0540}

\bibitem[{{Burrows} {et~al.}(1997){Burrows}, {Marley}, {Hubbard}, {Lunine},
  {Guillot}, {Saumon}, {Freedman}, {Sudarsky}, \& {Sharp}}]{Burrows+1997}
{Burrows}, A., {Marley}, M., {Hubbard}, W.~B., {et~al.} 1997, \apj, 491, 856,
  \dodoi{10.1086/305002}

\bibitem[{{Choi} {et~al.}(2016){Choi}, {Dotter}, {Conroy}, {Cantiello},
  {Paxton}, \& {Johnson}}]{mesa5}
{Choi}, J., {Dotter}, A., {Conroy}, C., {et~al.} 2016, \apj, 823, 102,
  \dodoi{10.3847/0004-637X/823/2/102}

\bibitem[{{Currie} {et~al.}(2012){Currie}, {Fukagawa}, {Thalmann}, {Matsumura},
  \& {Plavchan}}]{Currie+2012}
{Currie}, T., {Fukagawa}, M., {Thalmann}, C., {Matsumura}, S., \& {Plavchan},
  P. 2012, \apjl, 755, L34, \dodoi{10.1088/2041-8205/755/2/L34}

\bibitem[{{Currie} {et~al.}(2011){Currie}, {Burrows}, {Itoh}, {Matsumura},
  {Fukagawa}, {Apai}, {Madhusudhan}, {Hinz}, {Rodigas}, {Kasper}, {Pyo}, \&
  {Ogino}}]{Currie+2011}
{Currie}, T., {Burrows}, A., {Itoh}, Y., {et~al.} 2011, \apj, 729, 128,
  \dodoi{10.1088/0004-637X/729/2/128}

\bibitem[{{Currie} {et~al.}(2014){Currie}, {Burrows}, {Girard}, {Cloutier},
  {Fukagawa}, {Sorahana}, {Kuchner}, {Kenyon}, {Madhusudhan}, {Itoh},
  {Jayawardhana}, {Matsumura}, \& {Pyo}}]{Currie+2014}
{Currie}, T., {Burrows}, A., {Girard}, J.~H., {et~al.} 2014, \apj, 795, 133,
  \dodoi{10.1088/0004-637X/795/2/133}

\bibitem[{{De Rosa} {et~al.}(2020){De Rosa}, {Nguyen}, {Chilcote}, {Macintosh},
  {Perrin}, {Konopacky}, {Wang}, {Duch{\^e}ne}, {Nielsen}, {Rameau}, {Ammons},
  {Bailey}, {Barman}, {Bulger}, {Cotten}, {Doyon}, {Esposito}, {Fitzgerald},
  {Follette}, {Gerard}, {Goodsell}, {Graham}, {Greenbaum}, {Hibon}, {Hung},
  {Ingraham}, {Kalas}, {Larkin}, {Maire}, {Marchis}, {Marley}, {Marois},
  {Metchev}, {Millar-Blanchaer}, {Oppenheimer}, {Palmer}, {Patience},
  {Poyneer}, {Pueyo}, {Rajan}, {Rantakyr{\"o}}, {Ruffio}, {Savransky},
  {Schneider}, {Sivaramakrishnan}, {Song}, {Soummer}, {Thomas}, {Wallace},
  {Ward-Duong}, {Wiktorowicz}, \& {Wolff}}]{DeRosa+2020}
{De Rosa}, R.~J., {Nguyen}, M.~M., {Chilcote}, J., {et~al.} 2020, Journal of
  Astronomical Telescopes, Instruments, and Systems, 6, 015006,
  \dodoi{10.1117/1.JATIS.6.1.015006}

\bibitem[{{Dotter}(2016)}]{mesa4}
{Dotter}, A. 2016, \apjs, 222, 8, \dodoi{10.3847/0067-0049/222/1/8}

\bibitem[{{Doyon} {et~al.}(2010){Doyon}, {Lafreni{\`e}re}, {Artigau}, {Malo},
  \& {Marois}}]{Doyon2010}
{Doyon}, R., {Lafreni{\`e}re}, D., {Artigau}, E., {Malo}, L., \& {Marois}, C.
  2010, in In the Spirit of Lyot 2010, E42

\bibitem[{{Dupuy} {et~al.}(2019){Dupuy}, {Brandt}, {Kratter}, \&
  {Bowler}}]{Dupuy+2019}
{Dupuy}, T.~J., {Brandt}, T.~D., {Kratter}, K.~M., \& {Bowler}, B.~P. 2019,
  \apjl, 871, L4, \dodoi{10.3847/2041-8213/aafb31}

\bibitem[{{Esposito} {et~al.}(2013){Esposito}, {Mesa}, {Skemer}, {Arcidiacono},
  {Claudi}, {Desidera}, {Gratton}, {Mannucci}, {Marzari}, {Masciadri}, {Close},
  {Hinz}, {Kulesa}, {McCarthy}, {Males}, {Agapito}, {Argomedo}, {Boutsia},
  {Briguglio}, {Brusa}, {Busoni}, {Cresci}, {Fini}, {Fontana}, {Guerra},
  {Hill}, {Miller}, {Paris}, {Pinna}, {Puglisi}, {Quiros-Pacheco}, {Riccardi},
  {Stefanini}, {Testa}, {Xompero}, \& {Woodward}}]{esposito2013}
{Esposito}, S., {Mesa}, D., {Skemer}, A., {et~al.} 2013, \aap, 549, A52,
  \dodoi{10.1051/0004-6361/201219212}

\bibitem[{{Fabrycky} \& {Murray-Clay}(2010)}]{Fabrycky+Murray-Clay2010}
{Fabrycky}, D.~C., \& {Murray-Clay}, R.~A. 2010, \apj, 710, 1408,
  \dodoi{10.1088/0004-637X/710/2/1408}

\bibitem[{{Faramaz} {et~al.}(2021){Faramaz}, {Marino}, {Booth}, {Matr{\`a}},
  {Mamajek}, {Bryden}, {Stapelfeldt}, {Casassus}, {Cuadra}, {Hales}, \&
  {Zurlo}}]{Faramaz+2021}
{Faramaz}, V., {Marino}, S., {Booth}, M., {et~al.} 2021, \aj, 161, 271,
  \dodoi{10.3847/1538-3881/abf4e0}

\bibitem[{{Foreman-Mackey} {et~al.}(2013){Foreman-Mackey}, {Hogg}, {Lang}, \&
  {Goodman}}]{foremanmackey13}
{Foreman-Mackey}, D., {Hogg}, D.~W., {Lang}, D., \& {Goodman}, J. 2013, \pasp,
  125, 306, \dodoi{10.1086/670067}

\bibitem[{{Fortney} {et~al.}(2008){Fortney}, {Marley}, {Saumon}, \&
  {Lodders}}]{fortneyetal2008}
{Fortney}, J.~J., {Marley}, M.~S., {Saumon}, D., \& {Lodders}, K. 2008, \apj,
  683, 1104, \dodoi{10.1086/589942}

\bibitem[{{Gagn{\'e}} {et~al.}(2018){Gagn{\'e}}, {Mamajek}, {Malo}, {Riedel},
  {Rodriguez}, {Lafreni{\`e}re}, {Faherty}, {Roy-Loubier}, {Pueyo}, {Robin}, \&
  {Doyon}}]{BANYAN}
{Gagn{\'e}}, J., {Mamajek}, E.~E., {Malo}, L., {et~al.} 2018, \apj, 856, 23,
  \dodoi{10.3847/1538-4357/aaae09}

\bibitem[{{Gaia Collaboration} {et~al.}(2016){Gaia Collaboration}, {Prusti},
  {de Bruijne}, {Brown}, {Vallenari}, {Babusiaux}, {Bailer-Jones}, {Bastian},
  {Biermann}, {Evans}, \& et~al.}]{gaiamission}
{Gaia Collaboration}, {Prusti}, T., {de Bruijne}, J.~H.~J., {et~al.} 2016,
  \aap, 595, A1, \dodoi{10.1051/0004-6361/201629272}

\bibitem[{{Gaia Collaboration} {et~al.}(2021){Gaia Collaboration}, {Brown},
  {Vallenari}, {Prusti}, {de Bruijne}, {Babusiaux}, {Biermann}, {Creevey},
  {Evans}, {Eyer}, {Hutton}, {Jansen}, {Jordi}, {Klioner}, {Lammers},
  {Lindegren}, {Luri}, {Mignard}, {Panem}, {Pourbaix}, {Randich}, {Sartoretti},
  {Soubiran}, {Walton}, {Arenou}, {Bailer-Jones}, {Bastian}, {Cropper},
  {Drimmel}, {Katz}, {Lattanzi}, {van Leeuwen}, {Bakker}, {Cacciari},
  {Casta{\~n}eda}, {De Angeli}, {Ducourant}, {Fabricius}, {Fouesneau},
  {Fr{\'e}mat}, {Guerra}, {Guerrier}, {Guiraud}, {Jean-Antoine Piccolo},
  {Masana}, {Messineo}, {Mowlavi}, {Nicolas}, {Nienartowicz}, {Pailler},
  {Panuzzo}, {Riclet}, {Roux}, {Seabroke}, {Sordo}, {Tanga}, {Th{\'e}venin},
  {Gracia-Abril}, {Portell}, {Teyssier}, {Altmann}, {Andrae}, {Bellas-Velidis},
  {Benson}, {Berthier}, {Blomme}, {Brugaletta}, {Burgess}, {Busso}, {Carry},
  {Cellino}, {Cheek}, {Clementini}, {Damerdji}, {Davidson}, {Delchambre},
  {Dell'Oro}, {Fern{\'a}ndez-Hern{\'a}ndez}, {Galluccio}, {Garc{\'\i}a-Lario},
  {Garcia-Reinaldos}, {Gonz{\'a}lez-N{\'u}{\~n}ez}, {Gosset}, {Haigron},
  {Halbwachs}, {Hambly}, {Harrison}, {Hatzidimitriou}, {Heiter},
  {Hern{\'a}ndez}, {Hestroffer}, {Hodgkin}, {Holl}, {Jan{\ss}en}, {Jevardat de
  Fombelle}, {Jordan}, {Krone-Martins}, {Lanzafame}, {L{\"o}ffler}, {Lorca},
  {Manteiga}, {Marchal}, {Marrese}, {Moitinho}, {Mora}, {Muinonen}, {Osborne},
  {Pancino}, {Pauwels}, {Petit}, {Recio-Blanco}, {Richards}, {Riello},
  {Rimoldini}, {Robin}, {Roegiers}, {Rybizki}, {Sarro}, {Siopis}, {Smith},
  {Sozzetti}, {Ulla}, {Utrilla}, {van Leeuwen}, {van Reeven}, {Abbas}, {Abreu
  Aramburu}, {Accart}, {Aerts}, {Aguado}, {Ajaj}, {Altavilla}, {{\'A}lvarez},
  {{\'A}lvarez Cid-Fuentes}, {Alves}, {Anderson}, {Anglada Varela}, {Antoja},
  {Audard}, {Baines}, {Baker}, {Balaguer-N{\'u}{\~n}ez}, {Balbinot}, {Balog},
  {Barache}, {Barbato}, {Barros}, {Barstow}, {Bartolom{\'e}}, {Bassilana},
  {Bauchet}, {Baudesson-Stella}, {Becciani}, {Bellazzini}, {Bernet}, {Bertone},
  {Bianchi}, {Blanco-Cuaresma}, {Boch}, {Bombrun}, {Bossini}, {Bouquillon},
  {Bragaglia}, {Bramante}, {Breedt}, {Bressan}, {Brouillet}, {Bucciarelli},
  {Burlacu}, {Busonero}, {Butkevich}, {Buzzi}, {Caffau}, {Cancelliere},
  {C{\'a}novas}, {Cantat-Gaudin}, {Carballo}, {Carlucci}, {Carnerero},
  {Carrasco}, {Casamiquela}, {Castellani}, {Castro-Ginard}, {Castro Sampol},
  {Chaoul}, {Charlot}, {Chemin}, {Chiavassa}, {Cioni}, {Comoretto}, {Cooper},
  {Cornez}, {Cowell}, {Crifo}, {Crosta}, {Crowley}, {Dafonte}, {Dapergolas},
  {David}, {David}, {de Laverny}, {De Luise}, {De March}, {De Ridder}, {de
  Souza}, {de Teodoro}, {de Torres}, {del Peloso}, {del Pozo}, {Delbo},
  {Delgado}, {Delgado}, {Delisle}, {Di Matteo}, {Diakite}, {Diener},
  {Distefano}, {Dolding}, {Eappachen}, {Edvardsson}, {Enke}, {Esquej}, {Fabre},
  {Fabrizio}, {Faigler}, {Fedorets}, {Fernique}, {Fienga}, {Figueras},
  {Fouron}, {Fragkoudi}, {Fraile}, {Franke}, {Gai}, {Garabato},
  {Garcia-Gutierrez}, {Garc{\'\i}a-Torres}, {Garofalo}, {Gavras}, {Gerlach},
  {Geyer}, {Giacobbe}, {Gilmore}, {Girona}, {Giuffrida}, {Gomel}, {Gomez},
  {Gonzalez-Santamaria}, {Gonz{\'a}lez-Vidal}, {Granvik},
  {Guti{\'e}rrez-S{\'a}nchez}, {Guy}, {Hauser}, {Haywood}, {Helmi}, {Hidalgo},
  {Hilger}, {H{\l}adczuk}, {Hobbs}, {Holland}, {Huckle}, {Jasniewicz},
  {Jonker}, {Juaristi Campillo}, {Julbe}, {Karbevska}, {Kervella}, {Khanna},
  {Kochoska}, {Kontizas}, {Kordopatis}, {Korn}, {Kostrzewa-Rutkowska},
  {Kruszy{\'n}ska}, {Lambert}, {Lanza}, {Lasne}, {Le Campion}, {Le Fustec},
  {Lebreton}, {Lebzelter}, {Leccia}, {Leclerc}, {Lecoeur-Taibi}, {Liao},
  {Licata}, {Lindstr{\o}m}, {Lister}, {Livanou}, {Lobel}, {Madrero Pardo},
  {Managau}, {Mann}, {Marchant}, {Marconi}, {Marcos Santos}, {Marinoni},
  {Marocco}, {Marshall}, {Martin Polo}, {Mart{\'\i}n-Fleitas}, {Masip},
  {Massari}, {Mastrobuono-Battisti}, {Mazeh}, {McMillan}, {Messina},
  {Michalik}, {Millar}, {Mints}, {Molina}, {Molinaro}, {Moln{\'a}r},
  {Montegriffo}, {Mor}, {Morbidelli}, {Morel}, {Morris}, {Mulone}, {Munoz},
  {Muraveva}, {Murphy}, {Musella}, {Noval}, {Ord{\'e}novic}, {Orr{\`u}},
  {Osinde}, {Pagani}, {Pagano}, {Palaversa}, {Palicio}, {Panahi}, {Pawlak},
  {Pe{\~n}alosa Esteller}, {Penttil{\"a}}, {Piersimoni}, {Pineau}, {Plachy},
  {Plum}, {Poggio}, {Poretti}, {Poujoulet}, {Pr{\v{s}}a}, {Pulone}, {Racero},
  {Ragaini}, {Rainer}, {Raiteri}, {Rambaux}, {Ramos}, {Ramos-Lerate}, {Re
  Fiorentin}, {Regibo}, {Reyl{\'e}}, {Ripepi}, {Riva}, {Rixon}, {Robichon},
  {Robin}, {Roelens}, {Rohrbasser}, {Romero-G{\'o}mez}, {Rowell}, {Royer},
  {Rybicki}, {Sadowski}, {Sagrist{\`a} Sell{\'e}s}, {Sahlmann}, {Salgado},
  {Salguero}, {Samaras}, {Sanchez Gimenez}, {Sanna}, {Santove{\~n}a},
  {Sarasso}, {Schultheis}, {Sciacca}, {Segol}, {Segovia}, {S{\'e}gransan},
  {Semeux}, {Shahaf}, {Siddiqui}, {Siebert}, {Siltala}, {Slezak}, {Smart},
  {Solano}, {Solitro}, {Souami}, {Souchay}, {Spagna}, {Spoto}, {Steele},
  {Steidelm{\"u}ller}, {Stephenson}, {S{\"u}veges}, {Szabados}, {Szegedi-Elek},
  {Taris}, {Tauran}, {Taylor}, {Teixeira}, {Thuillot}, {Tonello}, {Torra},
  {Torra}, {Turon}, {Unger}, {Vaillant}, {van Dillen}, {Vanel}, {Vecchiato},
  {Viala}, {Vicente}, {Voutsinas}, {Weiler}, {Wevers}, {Wyrzykowski}, {Yoldas},
  {Yvard}, {Zhao}, {Zorec}, {Zucker}, {Zurbach}, \& {Zwitter}}]{GaiaEDR3}
{Gaia Collaboration}, {Brown}, A.~G.~A., {Vallenari}, A., {et~al.} 2021, \aap,
  649, A1, \dodoi{10.1051/0004-6361/202039657}

\bibitem[{{Galicher} {et~al.}(2011){Galicher}, {Marois}, {Macintosh}, {Barman},
  \& {Konopacky}}]{Galicher+2011}
{Galicher}, R., {Marois}, C., {Macintosh}, B., {Barman}, T., \& {Konopacky}, Q.
  2011, \apjl, 739, L41, \dodoi{10.1088/2041-8205/739/2/L41}

\bibitem[{{Goodman} \& {Weare}(2010)}]{goodmanWeare2010}
{Goodman}, J., \& {Weare}, J. 2010, Communications in Applied Mathematics and
  Computational Science, Vol.~5, No.~1, p.~65-80, 2010, 5, 65,
  \dodoi{10.2140/camcos.2010.5.65}

\bibitem[{{Go{\'z}dziewski} \& {Migaszewski}(2014)}]{gozdziewski2014}
{Go{\'z}dziewski}, K., \& {Migaszewski}, C. 2014, \mnras, 440, 3140,
  \dodoi{10.1093/mnras/stu455}

\bibitem[{{Go{\'z}dziewski} \&
  {Migaszewski}(2018)}]{Gozdziewski+Migaszewski2018}
---. 2018, \apjs, 238, 6, \dodoi{10.3847/1538-4365/aad3d3}

\bibitem[{{Go{\'z}dziewski} \&
  {Migaszewski}(2020)}]{Gozdziewski+Migaszewski2020}
---. 2020, \apjl, 902, L40, \dodoi{10.3847/2041-8213/abb881}

\bibitem[{{Gravity Collaboration} {et~al.}(2019){Gravity Collaboration},
  {Lacour}, {Nowak}, {Wang}, {Pfuhl}, {Eisenhauer}, {Abuter}, {Amorim},
  {Anugu}, {Benisty}, {Berger}, {Beust}, {Blind}, {Bonnefoy}, {Bonnet},
  {Bourget}, {Brandner}, {Buron}, {Collin}, {Charnay}, {Chapron}, {Cl{\'e}net},
  {Coud{\'e} Du Foresto}, {de Zeeuw}, {Deen}, {Dembet}, {Dexter}, {Duvert},
  {Eckart}, {F{\"o}rster Schreiber}, {F{\'e}dou}, {Garcia}, {Garcia Lopez},
  {Gao}, {Gendron}, {Genzel}, {Gillessen}, {Gordo}, {Greenbaum}, {Habibi},
  {Haubois}, {Hau{\ss}mann}, {Henning}, {Hippler}, {Horrobin}, {Hubert},
  {Jimenez Rosales}, {Jocou}, {Kendrew}, {Kervella}, {Kolb}, {Lagrange},
  {Lapeyr{\`e}re}, {Le Bouquin}, {L{\'e}na}, {Lippa}, {Lenzen}, {Maire},
  {Molli{\`e}re}, {Ott}, {Paumard}, {Perraut}, {Perrin}, {Pueyo}, {Rabien},
  {Ram{\'{\i}}rez}, {Rau}, {Rodr{\'{\i}}guez-Coira}, {Rousset},
  {Sanchez-Bermudez}, {Scheithauer}, {Schuhler}, {Straub}, {Straubmeier},
  {Sturm}, {Tacconi}, {Vincent}, {van Dishoeck}, {von Fellenberg}, {Wank},
  {Waisberg}, {Widmann}, {Wieprecht}, {Wiest}, {Wiezorrek}, {Woillez},
  {Yazici}, {Ziegler}, \& {Zins}}]{GRAVITY+2019}
{Gravity Collaboration}, {Lacour}, S., {Nowak}, M., {et~al.} 2019, \aap, 623,
  L11, \dodoi{10.1051/0004-6361/201935253}

\bibitem[{{Gravity Collaboration} {et~al.}(2020){Gravity Collaboration},
  {Nowak}, {Lacour}, {Molli{\`e}re}, {Wang}, {Charnay}, {van Dishoeck},
  {Abuter}, {Amorim}, {Berger}, {Beust}, {Bonnefoy}, {Bonnet}, {Brandner},
  {Buron}, {Cantalloube}, {Collin}, {Chapron}, {Cl{\'e}net}, {Coud{\'e} Du
  Foresto}, {de Zeeuw}, {Dembet}, {Dexter}, {Duvert}, {Eckart}, {Eisenhauer},
  {F{\"o}rster Schreiber}, {F{\'e}dou}, {Garcia Lopez}, {Gao}, {Gendron},
  {Genzel}, {Gillessen}, {Hau{\ss}mann}, {Henning}, {Hippler}, {Hubert},
  {Jocou}, {Kervella}, {Lagrange}, {Lapeyr{\`e}re}, {Le Bouquin}, {L{\'e}na},
  {Maire}, {Ott}, {Paumard}, {Paladini}, {Perraut}, {Perrin}, {Pueyo}, {Pfuhl},
  {Rabien}, {Rau}, {Rodr{\'\i}guez-Coira}, {Rousset}, {Scheithauer},
  {Shangguan}, {Straub}, {Straubmeier}, {Sturm}, {Tacconi}, {Vincent},
  {Widmann}, {Wieprecht}, {Wiezorrek}, {Woillez}, {Yazici}, \&
  {Ziegler}}]{Gravity+2020}
{Gravity Collaboration}, {Nowak}, M., {Lacour}, S., {et~al.} 2020, \aap, 633,
  A110, \dodoi{10.1051/0004-6361/201936898}

\bibitem[{{Gray} \& {Kaye}(1999)}]{grayandkaye1999}
{Gray}, R.~O., \& {Kaye}, A.~B. 1999, \aj, 118, 2993, \dodoi{10.1086/301134}

\bibitem[{{Greenbaum} {et~al.}(2018){Greenbaum}, {Pueyo}, {Ruffio}, {Wang}, {De
  Rosa}, {Aguilar}, {Rameau}, {Barman}, {Marois}, {Marley}, {Konopacky},
  {Rajan}, {Macintosh}, {Ansdell}, {Arriaga}, {Bailey}, {Bulger}, {Burrows},
  {Chilcote}, {Cotten}, {Doyon}, {Duch{\^e}ne}, {Fitzgerald}, {Follette},
  {Gerard}, {Goodsell}, {Graham}, {Hibon}, {Hung}, {Ingraham}, {Kalas},
  {Larkin}, {Maire}, {Marchis}, {Metchev}, {Millar-Blanchaer}, {Nielsen},
  {Norton}, {Oppenheimer}, {Palmer}, {Patience}, {Perrin}, {Poyneer},
  {Rantakyr{\"o}}, {Savransky}, {Schneider}, {Sivaramakrishnan}, {Song},
  {Soummer}, {Thomas}, {Wallace}, {Ward-Duong}, {Wiktorowicz}, \&
  {Wolff}}]{greenbaumetal2018}
{Greenbaum}, A.~Z., {Pueyo}, L., {Ruffio}, J.-B., {et~al.} 2018, \aj, 155, 226,
  \dodoi{10.3847/1538-3881/aabcb8}

\bibitem[{{Haffert} {et~al.}(2019){Haffert}, {Bohn}, {de Boer}, {Snellen},
  {Brinchmann}, {Girard}, {Keller}, \& {Bacon}}]{Haffert+2019}
{Haffert}, S.~Y., {Bohn}, A.~J., {de Boer}, J., {et~al.} 2019, Nature
  Astronomy, 3, 749, \dodoi{10.1038/s41550-019-0780-5}

\bibitem[{{Harris} {et~al.}(2020){Harris}, {Millman}, {van der Walt},
  {Gommers}, {Virtanen}, {Cournapeau}, {Wieser}, {Taylor}, {Berg}, {Smith},
  {Kern}, {Picus}, {Hoyer}, {van Kerkwijk}, {Brett}, {Haldane}, {del R{\'\i}o},
  {Wiebe}, {Peterson}, {G{\'e}rard-Marchant}, {Sheppard}, {Reddy}, {Weckesser},
  {Abbasi}, {Gohlke}, \& {Oliphant}}]{Harris+2020}
{Harris}, C.~R., {Millman}, K.~J., {van der Walt}, S.~J., {et~al.} 2020, \nat,
  585, 357, \dodoi{10.1038/s41586-020-2649-2}

\bibitem[{{Hillenbrand} \& {White}(2004)}]{hillenbrand2004}
{Hillenbrand}, L.~A., \& {White}, R.~J. 2004, \apj, 604, 741,
  \dodoi{10.1086/382021}

\bibitem[{{Hinz} {et~al.}(2010){Hinz}, {Rodigas}, {Kenworthy}, {Sivanandam},
  {Heinze}, {Mamajek}, \& {Meyer}}]{hinzetal2010}
{Hinz}, P.~M., {Rodigas}, T.~J., {Kenworthy}, M.~A., {et~al.} 2010, \apj, 716,
  417, \dodoi{10.1088/0004-637X/716/1/417}

\bibitem[{Hunter(2007)}]{matplotlib}
Hunter, J.~D. 2007, Computing in Science Engineering, 9, 90,
  \dodoi{10.1109/MCSE.2007.55}

\bibitem[{{Kaye} {et~al.}(1999){Kaye}, {Handler}, {Krisciunas}, {Poretti}, \&
  {Zerbi}}]{Kaye+1999}
{Kaye}, A.~B., {Handler}, G., {Krisciunas}, K., {Poretti}, E., \& {Zerbi},
  F.~M. 1999, \pasp, 111, 840, \dodoi{10.1086/316399}

\bibitem[{{Keppler} {et~al.}(2018){Keppler}, {Benisty}, {M{\"u}ller},
  {Henning}, {van Boekel}, {Cantalloube}, {Ginski}, {van Holstein}, {Maire},
  {Pohl}, {Samland }, {Avenhaus}, {Baudino}, {Boccaletti}, {de Boer},
  {Bonnefoy}, {Chauvin}, {Desidera}, {Langlois}, {Lazzoni}, {Marleau},
  {Mordasini}, {Pawellek}, {Stolker}, {Vigan}, {Zurlo}, {Birnstiel},
  {Brandner}, {Feldt}, {Flock}, {Girard}, {Gratton}, {Hagelberg}, {Isella},
  {Janson}, {Juhasz}, {Kemmer}, {Kral}, {Lagrange}, {Launhardt}, {Matter},
  {M{\'e}nard}, {Milli}, {Molli{\`e}re}, {Olofsson}, {P{\'e}rez}, {Pinilla},
  {Pinte}, {Quanz}, {Schmidt}, {Udry}, {Wahhaj}, {Williams}, {Buenzli},
  {Cudel}, {Dominik}, {Galicher}, {Kasper}, {Lannier}, {Mesa}, {Mouillet},
  {Peretti}, {Perrot}, {Salter}, {Sissa}, {Wildi}, {Abe}, {Antichi},
  {Augereau}, {Baruffolo}, {Baudoz}, {Bazzon}, {Beuzit}, {Blanchard}, {Brems},
  {Buey}, {De Caprio}, {Carbillet}, {Carle}, {Cascone}, {Cheetham}, {Claudi},
  {Costille}, {Delboulb{\'e}}, {Dohlen}, {Fantinel}, {Feautrier}, {Fusco},
  {Giro}, {Gluck}, {Gry}, {Hubin}, {Hugot}, {Jaquet}, {Le Mignant}, {Llored},
  {Madec}, {Magnard}, {Martinez}, {Maurel}, {Meyer}, {M{\"o}ller-Nilsson},
  {Moulin}, {Mugnier}, {Orign{\'e}}, {Pavlov}, {Perret}, {Petit}, {Pragt},
  {Puget}, {Rabou}, {Ramos}, {Rigal}, {Rochat}, {Roelfsema}, {Rousset}, {Roux},
  {Salasnich}, {Sauvage}, {Sevin}, {Soenke}, {Stadler}, {Suarez}, {Turatto}, \&
  {Weber}}]{Keppler+2018}
{Keppler}, M., {Benisty}, M., {M{\"u}ller}, A., {et~al.} 2018, \aap, 617, A44,
  \dodoi{10.1051/0004-6361/201832957}

\bibitem[{{Konopacky} \& {Barman}(2018)}]{2018haex}
{Konopacky}, Q.~M., \& {Barman}, T.~S. 2018, {HR8799: Imaging a System of
  Exoplanets}, ed. H.~J. {Deeg} \& J.~A. {Belmonte}, 36,
  \dodoi{10.1007/978-3-319-55333-7\_36}

\bibitem[{{Konopacky} {et~al.}(2016){Konopacky}, {Marois}, {Macintosh},
  {Galicher}, {Barman}, {Metchev}, \& {Zuckerman}}]{konopacky2016}
{Konopacky}, Q.~M., {Marois}, C., {Macintosh}, B.~A., {et~al.} 2016, \aj, 152,
  28, \dodoi{10.3847/0004-6256/152/2/28}

\bibitem[{{Lacour} {et~al.}(2021){Lacour}, {Wang}, {Rodet}, {Nowak},
  {Shangguan}, {Beust}, {Lagrange}, {Abuter}, {Amorim}, {Asensio-Torres},
  {Benisty}, {Berger}, {Blunt}, {Boccaletti}, {Bohn}, {Bolzer}, {Bonnefoy},
  {Bonnet}, {Bourdarot}, {Brandner}, {Cantalloube}, {Caselli}, {Charnay},
  {Chauvin}, {Choquet}, {Christiaens}, {Cl{\'e}net}, {Coud{\'e} Du Foresto},
  {Cridland}, {Dembet}, {Dexter}, {de Zeeuw}, {Drescher}, {Duvert}, {Eckart},
  {Eisenhauer}, {Gao}, {Garcia}, {Garcia Lopez}, {Gendron}, {Genzel},
  {Gillessen}, {Girard}, {Haubois}, {Hei{\ss}el}, {Henning}, {Hinkley},
  {Hippler}, {Horrobin}, {Houll{\'e}}, {Hubert}, {Jocou}, {Kammerer},
  {Keppler}, {Kervella}, {Kreidberg}, {Lapeyr{\`e}re}, {Le Bouquin},
  {L{\'e}na}, {Lutz}, {Maire}, {M{\'e}rand}, {Molli{\`e}re}, {Monnier},
  {Mouillet}, {Nasedkin}, {Ott}, {Otten}, {Paladini}, {Paumard}, {Perraut},
  {Perrin}, {Pfuhl}, {Rickman}, {Pueyo}, {Rameau}, {Rousset}, {Rustamkulov},
  {Samland}, {Shimizu}, {Sing}, {Stadler}, {Stolker}, {Straub}, {Straubmeier},
  {Sturm}, {Tacconi}, {van Dishoeck}, {Vigan}, {Vincent}, {von Fellenberg},
  {Ward-Duong}, {Widmann}, {Wieprecht}, {Wiezorrek}, {Woillez}, {Yazici},
  {Young}, \& {Gravity Collaboration}}]{Lacour+2021}
{Lacour}, S., {Wang}, J.~J., {Rodet}, L., {et~al.} 2021, \aap, 654, L2,
  \dodoi{10.1051/0004-6361/202141889}

\bibitem[{{Lafreni{\`e}re} {et~al.}(2009){Lafreni{\`e}re}, {Marois}, {Doyon},
  \& {Barman}}]{Lafreniere+2009}
{Lafreni{\`e}re}, D., {Marois}, C., {Doyon}, R., \& {Barman}, T. 2009, \apjl,
  694, L148, \dodoi{10.1088/0004-637X/694/2/L148}

\bibitem[{Lafreni{\`e}re {et~al.}(2007)Lafreni{\`e}re, Marois, Doyon, Nadeau,
  \& Artigau}]{Lafreniere:2007bg}
Lafreni{\`e}re, D., Marois, C., Doyon, R., Nadeau, D., \& Artigau, {\'E}. 2007,
  ApJ, 660, 770

\bibitem[{{Lagrange} {et~al.}(2009){Lagrange}, {Gratadour}, {Chauvin}, {Fusco},
  {Ehrenreich}, {Mouillet}, {Rousset}, {Rouan}, {Allard}, {Gendron}, {Charton},
  {Mugnier}, {Rabou}, {Montri}, \& {Lacombe}}]{Lagrange+2009}
{Lagrange}, A.~M., {Gratadour}, D., {Chauvin}, G., {et~al.} 2009, \aap, 493,
  L21, \dodoi{10.1051/0004-6361:200811325}

\bibitem[{{Lagrange} {et~al.}(2010){Lagrange}, {Bonnefoy}, {Chauvin}, {Apai},
  {Ehrenreich}, {Boccaletti}, {Gratadour}, {Rouan}, {Mouillet}, {Lacour}, \&
  {Kasper}}]{Lagrange+2010}
{Lagrange}, A.~M., {Bonnefoy}, M., {Chauvin}, G., {et~al.} 2010, Science, 329,
  57, \dodoi{10.1126/science.1187187}

\bibitem[{{Lagrange} {et~al.}(2019){Lagrange}, {Meunier}, {Rubini}, {Keppler},
  {Galland}, {Chapellier}, {Michel}, {Balona}, {Beust}, {Guillot}, {Grandjean},
  {Borgniet}, {M{\'e}karnia}, {Wilson}, {Kiefer}, {Bonnefoy}, {Lillo-Box},
  {Pantoja}, {Jones}, {Iglesias}, {Rodet}, {Diaz}, {Zapata}, {Abe}, \&
  {Schmider}}]{Lagrange+2019}
{Lagrange}, A.~M., {Meunier}, N., {Rubini}, P., {et~al.} 2019, Nature
  Astronomy, 3, 1135, \dodoi{10.1038/s41550-019-0857-1}

\bibitem[{{Lagrange} {et~al.}(2020){Lagrange}, {Rubini}, {Nowak}, {Lacour},
  {Grandjean}, {Boccaletti}, {Langlois}, {Delorme}, {Gratton}, {Wang},
  {Flasseur}, {Galicher}, {Kral}, {Meunier}, {Beust}, {Babusiaux}, {Le
  Coroller}, {Thebault}, {Kervella}, {Zurlo}, {Maire}, {Wahhaj}, {Amorim},
  {Asensio-Torres}, {Benisty}, {Berger}, {Bonnefoy}, {Brandner}, {Cantalloube},
  {Charnay}, {Chauvin}, {Choquet}, {Cl{\'e}net}, {Christiaens}, {Coud{\'e} Du
  Foresto}, {de Zeeuw}, {Desidera}, {Duvert}, {Eckart}, {Eisenhauer},
  {Galland}, {Gao}, {Garcia}, {Garcia Lopez}, {Gendron}, {Genzel}, {Gillessen},
  {Girard}, {Hagelberg}, {Haubois}, {Henning}, {Heissel}, {Hippler},
  {Horrobin}, {Janson}, {Kammerer}, {Kenworthy}, {Keppler}, {Kreidberg},
  {Lapeyr{\`e}re}, {Le Bouquin}, {L{\'e}na}, {M{\'e}rand}, {Messina},
  {Molli{\`e}re}, {Monnier}, {Ott}, {Otten}, {Paumard}, {Paladini}, {Perraut},
  {Perrin}, {Pueyo}, {Pfuhl}, {Rodet}, {Rodriguez-Coira}, {Rousset}, {Samland},
  {Shangguan}, {Schmidt}, {Straub}, {Straubmeier}, {Stolker}, {Vigan},
  {Vincent}, {Widmann}, {Woillez}, \& {Gravity Collaboration}}]{Lagrange+2020}
{Lagrange}, A.~M., {Rubini}, P., {Nowak}, M., {et~al.} 2020, \aap, 642, A18,
  \dodoi{10.1051/0004-6361/202038823}

\bibitem[{{Lindegren} {et~al.}(2021){Lindegren}, {Bastian}, {Biermann},
  {Bombrun}, {de Torres}, {Gerlach}, {Geyer}, {Hern{\'a}ndez}, {Hilger},
  {Hobbs}, {Klioner}, {Lammers}, {McMillan}, {Ramos-Lerate},
  {Steidelm{\"u}ller}, {Stephenson}, \& {van Leeuwen}}]{Lindegren+2021}
{Lindegren}, L., {Bastian}, U., {Biermann}, M., {et~al.} 2021, \aap, 649, A4,
  \dodoi{10.1051/0004-6361/202039653}

\bibitem[{{Liu}(2004)}]{Liu2004}
{Liu}, M.~C. 2004, Science, 305, 1442, \dodoi{10.1126/science.1102929}

\bibitem[{{Maire} {et~al.}(2015){Maire}, {Skemer}, {Hinz}, {Desidera},
  {Esposito}, {Gratton}, {Marzari}, {Skrutskie}, {Biller}, {Defr{\`e}re},
  {Bailey}, {Leisenring}, {Apai}, {Bonnefoy}, {Brandner}, {Buenzli}, {Claudi},
  {Close}, {Crepp}, {De Rosa}, {Eisner}, {Fortney}, {Henning}, {Hofmann},
  {Kopytova}, {Males}, {Mesa}, {Morzinski}, {Oza}, {Patience}, {Pinna},
  {Rajan}, {Schertl}, {Schlieder}, {Su}, {Vaz}, {Ward-Duong}, {Weigelt}, \&
  {Woodward}}]{mairetal2015}
{Maire}, A.-L., {Skemer}, A.~J., {Hinz}, P.~M., {et~al.} 2015, \aap, 576, A133,
  \dodoi{10.1051/0004-6361/201425185}

\bibitem[{{Malo} {et~al.}(2013){Malo}, {Doyon}, {Lafreni{\`e}re}, {Artigau},
  {Gagn{\'e}}, {Baron}, \& {Riedel}}]{maloetal2013}
{Malo}, L., {Doyon}, R., {Lafreni{\`e}re}, D., {et~al.} 2013, \apj, 762, 88,
  \dodoi{10.1088/0004-637X/762/2/88}

\bibitem[{{Marleau} \& {Cumming}(2014)}]{marleau&cumming2014}
{Marleau}, G.-D., \& {Cumming}, A. 2014, \mnras, 437, 1378,
  \dodoi{10.1093/mnras/stt1967}

\bibitem[{{Marley} {et~al.}(2007){Marley}, {Fortney}, {Hubickyj},
  {Bodenheimer}, \& {Lissauer}}]{marleyetal2007}
{Marley}, M.~S., {Fortney}, J.~J., {Hubickyj}, O., {Bodenheimer}, P., \&
  {Lissauer}, J.~J. 2007, \apj, 655, 541, \dodoi{10.1086/509759}

\bibitem[{Marois {et~al.}(2006)Marois, Lafreni{\`e}re, Doyon, Macintosh, \&
  Nadeau}]{Marois:2006df}
Marois, C., Lafreni{\`e}re, D., Doyon, R., Macintosh, B., \& Nadeau, D. 2006,
  ApJ, 641, 556

\bibitem[{{Marois} {et~al.}(2008){Marois}, {Macintosh}, {Barman}, {Zuckerman},
  {Song}, {Patience}, {Lafreni{\`e}re}, \& {Doyon}}]{marios2008}
{Marois}, C., {Macintosh}, B., {Barman}, T., {et~al.} 2008, Science, 322, 1348,
  \dodoi{10.1126/science.1166585}

\bibitem[{{Marois} {et~al.}(2010){Marois}, {Zuckerman}, {Konopacky},
  {Macintosh}, \& {Barman}}]{marios2010}
{Marois}, C., {Zuckerman}, B., {Konopacky}, Q.~M., {Macintosh}, B., \&
  {Barman}, T. 2010, \nat, 468, 1080, \dodoi{10.1038/nature09684}

\bibitem[{{Mesa} {et~al.}(2019){Mesa}, {Keppler}, {Cantalloube}, {Rodet},
  {Charnay}, {Gratton}, {Langlois}, {Boccaletti}, {Bonnefoy}, {Vigan},
  {Flasseur}, {Bae}, {Benisty}, {Chauvin}, {de Boer}, {Desidera}, {Henning},
  {Lagrange}, {Meyer}, {Milli}, {M{\"u}ller}, {Pairet}, {Zurlo}, {Antoniucci},
  {Baudino}, {Brown Sevilla}, {Cascone}, {Cheetham}, {Claudi}, {Delorme},
  {D'Orazi}, {Feldt}, {Hagelberg}, {Janson}, {Kral}, {Lagadec}, {Lazzoni},
  {Ligi}, {Maire}, {Martinez}, {Menard}, {Meunier}, {Perrot}, {Petrus},
  {Pinte}, {Rickman}, {Rochat}, {Rouan}, {Samland}, {Sauvage}, {Schmidt},
  {Udry}, {Weber}, \& {Wildi}}]{Mesa+2019}
{Mesa}, D., {Keppler}, M., {Cantalloube}, F., {et~al.} 2019, \aap, 632, A25,
  \dodoi{10.1051/0004-6361/201936764}

\bibitem[{{Metchev} {et~al.}(2009){Metchev}, {Marois}, \&
  {Zuckerman}}]{Metchev+2009}
{Metchev}, S., {Marois}, C., \& {Zuckerman}, B. 2009, \apjl, 705, L204,
  \dodoi{10.1088/0004-637X/705/2/L204}

\bibitem[{{Moya} {et~al.}(2010{\natexlab{a}}){Moya}, {Amado}, {Barrado},
  {Garc{\'{\i}}a Hern{\'a}ndez}, {Aberasturi}, {Montesinos}, \&
  {Aceituno}}]{moyaetal2010a}
{Moya}, A., {Amado}, P.~J., {Barrado}, D., {et~al.} 2010{\natexlab{a}}, \mnras,
  405, L81, \dodoi{10.1111/j.1745-3933.2010.00863.x}

\bibitem[{{Moya} {et~al.}(2010{\natexlab{b}}){Moya}, {Amado}, {Barrado},
  {Hern{\'a}ndez}, {Aberasturi}, {Montesinos}, \& {Aceituno}}]{moya2010b}
---. 2010{\natexlab{b}}, \mnras, 406, 566,
  \dodoi{10.1111/j.1365-2966.2010.16699.x}

\bibitem[{{Murphy} {et~al.}(2021){Murphy}, {Joyce}, {Bedding}, {White}, \&
  {Kama}}]{Murphy+2021}
{Murphy}, S.~J., {Joyce}, M., {Bedding}, T.~R., {White}, T.~R., \& {Kama}, M.
  2021, \mnras, 502, 1633, \dodoi{10.1093/mnras/stab144}

\bibitem[{{Murphy} {et~al.}(2015){Murphy}, {Corbally}, {Gray}, {Cheng}, {Neff},
  {Koen}, {Kuehn}, {Newsome}, \& {Riggs}}]{Murphy+2015}
{Murphy}, S.~J., {Corbally}, C.~J., {Gray}, R.~O., {et~al.} 2015, \pasa, 32,
  e036, \dodoi{10.1017/pasa.2015.34}

\bibitem[{Nelder \& Mead(1965)}]{Nelder:1965tk}
Nelder, J.~A., \& Mead, R. 1965, The Computer Journal, 7, 308

\bibitem[{{Nielsen} {et~al.}(2014){Nielsen}, {Liu}, {Wahhaj}, {Biller},
  {Hayward}, {Males}, {Close}, {Morzinski}, {Skemer}, {Kuchner}, {Rodigas},
  {Hinz}, {Chun}, {Ftaclas}, \& {Toomey}}]{Nielsen2014}
{Nielsen}, E.~L., {Liu}, M.~C., {Wahhaj}, Z., {et~al.} 2014, \apj, 794, 158,
  \dodoi{10.1088/0004-637X/794/2/158}

\bibitem[{{Nielsen} {et~al.}(2020){Nielsen}, {De Rosa}, {Wang}, {Sahlmann},
  {Kalas}, {Duch{\^e}ne}, {Rameau}, {Marley}, {Saumon}, {Macintosh},
  {Millar-Blanchaer}, {Nguyen}, {Ammons}, {Bailey}, {Barman}, {Bulger},
  {Chilcote}, {Cotten}, {Doyon}, {Esposito}, {Fitzgerald}, {Follette},
  {Gerard}, {Goodsell}, {Graham}, {Greenbaum}, {Hibon}, {Hung}, {Ingraham},
  {Konopacky}, {Larkin}, {Maire}, {Marchis}, {Marois}, {Metchev},
  {Oppenheimer}, {Palmer}, {Patience}, {Perrin}, {Poyneer}, {Pueyo}, {Rajan},
  {Rantakyr{\"o}}, {Ruffio}, {Savransky}, {Schneider}, {Sivaramakrishnan},
  {Song}, {Soummer}, {Thomas}, {Wallace}, {Ward-Duong}, {Wiktorowicz}, \&
  {Wolff}}]{Nielsen+2020}
{Nielsen}, E.~L., {De Rosa}, R.~J., {Wang}, J.~J., {et~al.} 2020, \aj, 159, 71,
  \dodoi{10.3847/1538-3881/ab5b92}

\bibitem[{{Nowak} {et~al.}(2020){Nowak}, {Lacour}, {Lagrange}, {Rubini},
  {Wang}, {Stolker}, {Abuter}, {Amorim}, {Asensio-Torres}, {Baub{\"o}ck},
  {Benisty}, {Berger}, {Beust}, {Blunt}, {Boccaletti}, {Bonnefoy}, {Bonnet},
  {Brandner}, {Cantalloube}, {Charnay}, {Choquet}, {Christiaens}, {Cl{\'e}net},
  {Coud{\'e} Du Foresto}, {Cridland}, {de Zeeuw}, {Dembet}, {Dexter},
  {Drescher}, {Duvert}, {Eckart}, {Eisenhauer}, {Gao}, {Garcia}, {Garcia
  Lopez}, {Gardner}, {Gendron}, {Genzel}, {Gillessen}, {Girard}, {Grandjean},
  {Haubois}, {Hei{\ss}el}, {Henning}, {Hinkley}, {Hippler}, {Horrobin},
  {Houll{\'e}}, {Hubert}, {Jim{\'e}nez-Rosales}, {Jocou}, {Kammerer},
  {Kervella}, {Keppler}, {Kreidberg}, {Kulikauskas}, {Lapeyr{\`e}re}, {Le
  Bouquin}, {L{\'e}na}, {M{\'e}rand}, {Maire}, {Molli{\`e}re}, {Monnier},
  {Mouillet}, {M{\"u}ller}, {Nasedkin}, {Ott}, {Otten}, {Paumard}, {Paladini},
  {Perraut}, {Perrin}, {Pueyo}, {Pfuhl}, {Rameau}, {Rodet},
  {Rodr{\'\i}guez-Coira}, {Rousset}, {Scheithauer}, {Shangguan}, {Stadler},
  {Straub}, {Straubmeier}, {Sturm}, {Tacconi}, {van Dishoeck}, {Vigan},
  {Vincent}, {von Fellenberg}, {Ward-Duong}, {Widmann}, {Wieprecht},
  {Wiezorrek}, {Woillez}, \& {Gravity Collaboration}}]{Nowak+2020}
{Nowak}, M., {Lacour}, S., {Lagrange}, A.~M., {et~al.} 2020, \aap, 642, L2,
  \dodoi{10.1051/0004-6361/202039039}

\bibitem[{{Paxton} {et~al.}(2011){Paxton}, {Bildsten}, {Dotter}, {Herwig},
  {Lesaffre}, \& {Timmes}}]{mesa1}
{Paxton}, B., {Bildsten}, L., {Dotter}, A., {et~al.} 2011, \apjs, 192, 3,
  \dodoi{10.1088/0067-0049/192/1/3}

\bibitem[{{Paxton} {et~al.}(2013){Paxton}, {Cantiello}, {Arras}, {Bildsten},
  {Brown}, {Dotter}, {Mankovich}, {Montgomery}, {Stello}, {Timmes}, \&
  {Townsend}}]{mesa2}
{Paxton}, B., {Cantiello}, M., {Arras}, P., {et~al.} 2013, \apjs, 208, 4,
  \dodoi{10.1088/0067-0049/208/1/4}

\bibitem[{{Paxton} {et~al.}(2015){Paxton}, {Marchant}, {Schwab}, {Bauer},
  {Bildsten}, {Cantiello}, {Dessart}, {Farmer}, {Hu}, {Langer}, {Townsend},
  {Townsley}, \& {Timmes}}]{mesa3}
{Paxton}, B., {Marchant}, P., {Schwab}, J., {et~al.} 2015, \apjs, 220, 15,
  \dodoi{10.1088/0067-0049/220/1/15}

\bibitem[{{Paxton} {et~al.}(2018){Paxton}, {Schwab}, {Bauer}, {Bildsten},
  {Blinnikov}, {Duffell}, {Farmer}, {Goldberg}, {Marchant}, {Sorokina},
  {Thoul}, {Townsend}, \& {Timmes}}]{mesa3.5}
{Paxton}, B., {Schwab}, J., {Bauer}, E.~B., {et~al.} 2018, \apjs, 234, 34,
  \dodoi{10.3847/1538-4365/aaa5a8}

\bibitem[{{Pueyo} {et~al.}(2015){Pueyo}, {Soummer}, {Hoffmann}, {Oppenheimer},
  {Graham}, {Zimmerman}, {Zhai}, {Wallace}, {Vescelus}, {Veicht}, {Vasisht},
  {Truong}, {Sivaramakrishnan}, {Shao}, {Roberts}, {Roberts}, {Rice}, {Parry},
  {Nilsson}, {Lockhart}, {Ligon}, {King}, {Hinkley}, {Hillenbrand}, {Hale},
  {Dekany}, {Crepp}, {Cady}, {Burruss}, {Brenner}, {Beichman}, \&
  {Baranec}}]{pueyo2015}
{Pueyo}, L., {Soummer}, R., {Hoffmann}, J., {et~al.} 2015, \apj, 803, 31,
  \dodoi{10.1088/0004-637X/803/1/31}

\bibitem[{{Rajan} {et~al.}(2015){Rajan}, {Barman}, {Soummer}, {Hagan},
  {Patience}, {Pueyo}, {Choquet}, {Konopacky}, {Macintosh}, \&
  {Marois}}]{Rajan+2015}
{Rajan}, A., {Barman}, T., {Soummer}, R., {et~al.} 2015, \apjl, 809, L33,
  \dodoi{10.1088/2041-8205/809/2/L33}

\bibitem[{{Rodriguez} \& {Zerbi}(1995)}]{Rodriguez&Zerbi1995}
{Rodriguez}, E., \& {Zerbi}, F.~M. 1995, Information Bulletin on Variable
  Stars, 4170, 1

\bibitem[{{Ruffio} {et~al.}(2019){Ruffio}, {Macintosh}, {Konopacky}, {Barman},
  {De Rosa}, {Wang}, {Wilcomb}, {Czekala}, \& {Marois}}]{Ruffio+2019}
{Ruffio}, J.-B., {Macintosh}, B., {Konopacky}, Q.~M., {et~al.} 2019, \aj, 158,
  200, \dodoi{10.3847/1538-3881/ab4594}

\bibitem[{{Ruffio} {et~al.}(2021){Ruffio}, {Konopacky}, {Barman}, {Macintosh},
  {Wilcomb}, {De Rosa}, {Wang}, {Czekala}, \& {Marois}}]{Ruffio+2021}
{Ruffio}, J.-B., {Konopacky}, Q.~M., {Barman}, T., {et~al.} 2021, arXiv
  e-prints, arXiv:2109.07614.
\newblock \doarXiv{2109.07614}

\bibitem[{{Sadakane}(2006)}]{Sadakane2006}
{Sadakane}, K. 2006, \pasj, 58, 1023, \dodoi{10.1093/pasj/58.6.1023}

\bibitem[{{Serenelli} {et~al.}(2021){Serenelli}, {Weiss}, {Aerts}, {Angelou},
  {Baroch}, {Bastian}, {Beck}, {Bergemann}, {Bestenlehner}, {Czekala},
  {Elias-Rosa}, {Escorza}, {Van Eylen}, {Feuillet}, {Gandolfi}, {Gieles},
  {Girardi}, {Lebreton}, {Lodieu}, {Martig}, {Miller Bertolami}, {Mombarg},
  {Morales}, {Moya}, {Nsamba}, {Pavlovski}, {Pedersen}, {Ribas}, {Schneider},
  {Silva Aguirre}, {Stassun}, {Tolstoy}, {Tremblay}, \&
  {Zwintz}}]{Serenelli+2020}
{Serenelli}, A., {Weiss}, A., {Aerts}, C., {et~al.} 2021, \aapr, 29, 4,
  \dodoi{10.1007/s00159-021-00132-9}

\bibitem[{Service {et~al.}(2016)Service, Lu, Campbell, Sitarski, Ghez, \&
  Anderson}]{Service:2016gk}
Service, M., Lu, J.~R., Campbell, R., {et~al.} 2016, PASP, 128, 095004

\bibitem[{{S{\'o}dor} {et~al.}(2014){S{\'o}dor}, {Chen{\'e}}, {De Cat},
  {Bogn{\'a}r}, {Wright}, {Marois}, {Walker}, {Matthews}, {Kallinger}, {Rowe},
  {Kuschnig}, {Guenther}, {Moffat}, {Rucinski}, {Sasselov}, \&
  {Weiss}}]{Sodor+2014}
{S{\'o}dor}, {\'A}., {Chen{\'e}}, A.~N., {De Cat}, P., {et~al.} 2014, \aap,
  568, A106, \dodoi{10.1051/0004-6361/201423976}

\bibitem[{{Soummer} {et~al.}(2011){Soummer}, {Hagan}, {Pueyo}, {Thormann},
  {Rajan}, \& {Marois}}]{soummer2011}
{Soummer}, R., {Hagan}, J.~B., {Pueyo}, L., {et~al.} 2011, \apj, 741, 55,
  \dodoi{10.1088/0004-637X/741/1/55}

\bibitem[{{Su} {et~al.}(2009){Su}, {Rieke}, {Stapelfeldt}, {Malhotra},
  {Bryden}, {Smith}, {Misselt}, {Moro-Martin}, \& {Williams}}]{Su+2009}
{Su}, K.~Y.~L., {Rieke}, G.~H., {Stapelfeldt}, K.~R., {et~al.} 2009, \apj, 705,
  314, \dodoi{10.1088/0004-637X/705/1/314}

\bibitem[{{van der Walt} {et~al.}(2011){van der Walt}, {Colbert}, \&
  {Varoquaux}}]{numpy}
{van der Walt}, S., {Colbert}, S.~C., \& {Varoquaux}, G. 2011, Computing in
  Science and Engineering, 13, 22, \dodoi{10.1109/MCSE.2011.37}

\bibitem[{{van Leeuwen}(2007)}]{vanLeeuwen2007}
{van Leeuwen}, F. 2007, \aap, 474, 653, \dodoi{10.1051/0004-6361:20078357}

\bibitem[{{Vandal} {et~al.}(2020){Vandal}, {Rameau}, \& {Doyon}}]{Vandal+2020}
{Vandal}, T., {Rameau}, J., \& {Doyon}, R. 2020, \aj, 160, 243,
  \dodoi{10.3847/1538-3881/abba30}

\bibitem[{{Veras} \& {Hinkley}(2021)}]{Veras+Hinkley2021}
{Veras}, D., \& {Hinkley}, S. 2021, \mnras, 505, 1557,
  \dodoi{10.1093/mnras/stab1311}

\bibitem[{{Vousden} {et~al.}(2016){Vousden}, {Farr}, \& {Mandel}}]{ptemcee}
{Vousden}, W.~D., {Farr}, W.~M., \& {Mandel}, I. 2016, \mnras, 455, 1919,
  \dodoi{10.1093/mnras/stv2422}

\bibitem[{{Wahhaj} {et~al.}(2021){Wahhaj}, {Milli}, {Romero}, {Cieza}, {Zurlo},
  {Vigan}, {Pe{\~n}a}, {Valdes}, {Cantalloube}, {Girard}, \&
  {Pantoja}}]{Wahhaj+2021}
{Wahhaj}, Z., {Milli}, J., {Romero}, C., {et~al.} 2021, \aap, 648, A26,
  \dodoi{10.1051/0004-6361/202038794}

\bibitem[{{Wang} {et~al.}(2018{\natexlab{a}}){Wang}, {Mawet}, {Fortney},
  {Hood}, {Morley}, \& {Benneke}}]{Wang+2018RV}
{Wang}, J., {Mawet}, D., {Fortney}, J.~J., {et~al.} 2018{\natexlab{a}}, \aj,
  156, 272, \dodoi{10.3847/1538-3881/aae47b}

\bibitem[{{Wang} {et~al.}(2016){Wang}, {Graham}, {Pueyo}, {Kalas},
  {Millar-Blanchaer}, {Ruffio}, {De Rosa}, {Ammons}, {Arriaga}, {Bailey},
  {Barman}, {Bulger}, {Burrows}, {Cardwell}, {Chen}, {Chilcote}, {Cotten},
  {Fitzgerald}, {Follette}, {Doyon}, {Duch{\^e}ne}, {Greenbaum}, {Hibon},
  {Hung}, {Ingraham}, {Konopacky}, {Larkin}, {Macintosh}, {Maire}, {Marchis},
  {Marley}, {Marois}, {Metchev}, {Nielsen}, {Oppenheimer}, {Palmer}, {Patel},
  {Patience}, {Perrin}, {Poyneer}, {Rajan}, {Rameau}, {Rantakyr{\"o}},
  {Savransky}, {Sivaramakrishnan}, {Song}, {Soummer}, {Thomas}, {Vasisht},
  {Vega}, {Wallace}, {Ward-Duong}, {Wiktorowicz}, \& {Wolff}}]{Wang2016}
{Wang}, J.~J., {Graham}, J.~R., {Pueyo}, L., {et~al.} 2016, \aj, 152, 97,
  \dodoi{10.3847/0004-6256/152/4/97}

\bibitem[{{Wang} {et~al.}(2018{\natexlab{b}}){Wang}, {Graham}, {Dawson},
  {Fabrycky}, {De Rosa}, {Pueyo}, {Konopacky}, {Macintosh}, {Marois}, {Chiang},
  {Ammons}, {Arriaga}, {Bailey}, {Barman}, {Bulger}, {Chilcote}, {Cotten},
  {Doyon}, {Duch{\^e}ne}, {Esposito}, {Fitzgerald}, {Follette}, {Gerard},
  {Goodsell}, {Greenbaum}, {Hibon}, {Hung}, {Ingraham}, {Kalas}, {Larkin},
  {Maire}, {Marchis}, {Marley}, {Metchev}, {Millar-Blanchaer}, {Nielsen},
  {Oppenheimer}, {Palmer}, {Patience}, {Perrin}, {Poyneer}, {Rajan}, {Rameau},
  {Rantakyr{\"o}}, {Ruffio}, {Savransky}, {Schneider}, {Sivaramakrishnan},
  {Song}, {Soummer}, {Thomas}, {Wallace}, {Ward-Duong}, {Wiktorowicz}, \&
  {Wolff}}]{Wang+2018}
{Wang}, J.~J., {Graham}, J.~R., {Dawson}, R., {et~al.} 2018{\natexlab{b}}, \aj,
  156, 192, \dodoi{10.3847/1538-3881/aae150}

\bibitem[{{Wang} {et~al.}(2021){Wang}, {Ruffio}, {Morris}, {Delorme},
  {Jovanovic}, {Pezzato}, {Echeverri}, {Finnerty}, {Hood}, {Zanazzi}, {Bryan},
  {Bond}, {Cetre}, {Martin}, {Mawet}, {Skemer}, {Baker}, {Xuan}, {Wallace},
  {Wang}, {Bartos}, {Blake}, {Boden}, {Buzard}, {Calvin}, {Chun}, {Doppmann},
  {Dupuy}, {Duch{\^e}ne}, {Feng}, {Fitzgerald}, {Fortney}, {Freedman},
  {Knutson}, {Konopacky}, {Lilley}, {Liu}, {Lopez}, {Lupu}, {Marley},
  {Meshkat}, {Miles}, {Millar-Blanchaer}, {Ragland}, {Roy}, {Ruane}, {Sappey},
  {Schofield}, {Weiss}, {Wetherell}, {Wizinowich}, \& {Ygouf}}]{Wang+2021}
{Wang}, J.~J., {Ruffio}, J.-B., {Morris}, E., {et~al.} 2021, \aj, 162, 148,
  \dodoi{10.3847/1538-3881/ac1349}

\bibitem[{{Wertz} {et~al.}(2017){Wertz}, {Absil}, {G{\'o}mez Gonz{\'a}lez},
  {Milli}, {Girard}, {Mawet}, \& {Pueyo}}]{Wertz+2017}
{Wertz}, O., {Absil}, O., {G{\'o}mez Gonz{\'a}lez}, C.~A., {et~al.} 2017, \aap,
  598, A83, \dodoi{10.1051/0004-6361/201628730}

\bibitem[{{Wilner} {et~al.}(2018){Wilner}, {MacGregor}, {Andrews}, {Hughes},
  {Matthews}, \& {Su}}]{wilner2018}
{Wilner}, D.~J., {MacGregor}, M.~A., {Andrews}, S.~M., {et~al.} 2018, \apj,
  855, 56, \dodoi{10.3847/1538-4357/aaacd7}

\bibitem[{Wizinowich(2013)}]{Wizinowich:2013dz}
Wizinowich, P. 2013, PASP, 125, 798

\bibitem[{{Wright} {et~al.}(2011){Wright}, {Chen{\'e}}, {De Cat}, {Marois},
  {Mathias}, {Macintosh}, {Isaacs}, {Lehmann}, \& {Hartmann}}]{Wright+2011}
{Wright}, D.~J., {Chen{\'e}}, A.~N., {De Cat}, P., {et~al.} 2011, \apjl, 728,
  L20, \dodoi{10.1088/2041-8205/728/1/L20}

\bibitem[{{Yi} {et~al.}(2001){Yi}, {Demarque}, {Kim}, {Lee}, {Ree}, {Lejeune},
  \& {Barnes}}]{Yi+2001}
{Yi}, S., {Demarque}, P., {Kim}, Y.-C., {et~al.} 2001, \apjs, 136, 417,
  \dodoi{10.1086/321795}

\bibitem[{{Zerbi} {et~al.}(1999){Zerbi}, {Rodr{\'{\i}}guez}, {Garrido},
  {Mart{\'{\i}}n}, {Arellano Ferro}, {Sareyan}, {Krisciunas}, {Akan}, {Evren},
  {Ibano{\v g}lu}, {Keskin}, {Pekunlu}, {Tunca}, {Luedeke}, {Paparo}, {Nuspl},
  \& {Guerrero}}]{zerbi1999}
{Zerbi}, F.~M., {Rodr{\'{\i}}guez}, E., {Garrido}, R., {et~al.} 1999, \mnras,
  303, 275, \dodoi{10.1046/j.1365-8711.1999.02209.x}

\bibitem[{{Zuckerman} {et~al.}(2011){Zuckerman}, {Rhee}, {Song}, \&
  {Bessell}}]{zuckerman2011}
{Zuckerman}, B., {Rhee}, J.~H., {Song}, I., \& {Bessell}, M.~S. 2011, \apj,
  732, 61, \dodoi{10.1088/0004-637X/732/2/61}

\bibitem[{{Zurlo} {et~al.}(2016){Zurlo}, {Vigan}, {Galicher}, {Maire}, {Mesa},
  {Gratton}, {Chauvin}, {Kasper}, {Moutou}, {Bonnefoy}, {Desidera}, {Abe},
  {Apai}, {Baruffolo}, {Baudoz}, {Baudrand}, {Beuzit}, {Blancard},
  {Boccaletti}, {Cantalloube}, {Carle}, {Cascone}, {Charton}, {Claudi},
  {Costille}, {de Caprio}, {Dohlen}, {Dominik}, {Fantinel}, {Feautrier},
  {Feldt}, {Fusco}, {Gigan}, {Girard}, {Gisler}, {Gluck}, {Gry}, {Henning},
  {Hugot}, {Janson}, {Jaquet}, {Lagrange}, {Langlois}, {Llored}, {Madec},
  {Magnard}, {Martinez}, {Maurel}, {Mawet}, {Meyer}, {Milli},
  {Moeller-Nilsson}, {Mouillet}, {Orign{\'e}}, {Pavlov}, {Petit}, {Puget},
  {Quanz}, {Rabou}, {Ramos}, {Rousset}, {Roux}, {Salasnich}, {Salter},
  {Sauvage}, {Schmid}, {Soenke}, {Stadler}, {Suarez}, {Turatto}, {Udry},
  {Vakili}, {Wahhaj}, {Wildi}, \& {Antichi}}]{zurloetal2016}
{Zurlo}, A., {Vigan}, A., {Galicher}, R., {et~al.} 2016, \aap, 587, A57,
  \dodoi{10.1051/0004-6361/201526835}

\end{thebibliography}
\end{document}